\documentclass[aps,prb,reprint,superscriptaddress,footinbib,longbibliography]{revtex4-2}

\usepackage{graphicx}
\usepackage{subfigure}
\usepackage{amsmath}
\usepackage{amssymb}
\usepackage{siunitx}                  
\usepackage{bm}                       
\usepackage{braket}                   
\usepackage{hyperref}

\renewcommand{\d}[0]{\mathrm{d}}      
\renewcommand{\Re}{\operatorname{Re}} 
\newcommand{\Tr}{\operatorname{Tr}}   
\newcommand{\adj}{\operatorname{adj}} 

\begin{document}

\title{
    Analytical Floquet Quantum Statistics from Nonequilibrium Green's Functions
}


\author{Yuhua Ren} 
\email{yuhua.ren@u.nus.edu}
\affiliation{Department of Physics, 
National University of Singapore, 
Singapore 117551, 
Singapore}

\author{Gaomin Tang}
\affiliation{Graduate School of China Academy of Engineering Physics, 
Beijing 100193, 
China}

\author{Hui Pan}
\affiliation{Department of Physics, 
National University of Singapore, 
Singapore 117551, 
Singapore}

\author{Jian-Sheng Wang}
\affiliation{Department of Physics, 
National University of Singapore, 
Singapore 117551, 
Singapore}

\bigskip

\begin{abstract}
    We derive an analytical expression for the steady-state quantum statistics of periodically driven quantum systems coupled to a bath using the nonequilibrium Green's function (NEGF) formalism. 
    By embedding Floquet theory into NEGF, we obtain closed expressions for the retarded, advanced, and lesser Green's functions in the Floquet representation, yielding the Floquet Fermi distribution in which the steady-state occupation is expressed as a weighted sum of Fermi functions evaluated at energies shifted by integer multiples of the driving frequency. 
    The weights are determined solely by the Fourier components of the micromotion operator, providing a transparent interpretation of Floquet sideband occupations. 
    We extend the analysis beyond diagonal commuting Hamiltonians treated previously, and derive the off‑diagonal decoherences that arise under degeneracies. 
    We further show that the robust Floquet distribution remains valid for a broad class of weakly coupled bath spectral functions beyond the ideal featureless-bath approximation. 
    Finally, we establish a Floquet version of the Landauer formula for the DC part of the current, in which the equilibrium Fermi functions are replaced by their Floquet-modified counterparts. 
    Together, these results provide a coherent description of Floquet quantum statistics and transport in periodically driven open quantum systems.
\end{abstract}

\maketitle

\section{Introduction}

Periodic driving has emerged as a powerful tool for controlling and engineering quantum systems far from equilibrium. 
Through Floquet engineering, time-periodic modulation can generate effective Hamiltonians with properties inaccessible in static systems, enabling ``on-demand'' realization of novel topological phases~\cite{oka_photovoltaic_2009,lindner_floquet_2011,wang_observation_2013}, dynamical stabilization~\cite{bukov_universal_2015,jiao_floquet_2026}, synthetic gauge fields~\cite{wang_floquet_2024}, and tunable band structures~\cite{oka_photovoltaic_2009,lindner_floquet_2011,rudner_band_2020,castro_floquet_2022}.
These ideas have found applications across a broad range of platforms, including driven semiconductors~\cite{sie_valley-selective_2015,esin_floquet_2020}, superconducting circuits~\cite{deng_observation_2015,zhao_probing_2022}, ultracold atoms~\cite{jotzu_experimental_2014,meinert_floquet_2016,miller_two-axis_2024}, and photonic systems~\cite{rechtsman_photonic_2013,maczewsky_observation_2017}. 
As experimental capabilities continue to advance, periodically driven quantum matter has become a central theme in modern condensed matter physics.

However, a realistically driven system is never perfectly isolated. 
Coupling to an external environment leads to dissipation, decoherence, and energy exchange, which fundamentally influence the system's steady-state properties. 
Understanding the interplay between periodic driving and bath-induced relaxation is therefore essential for connecting Floquet theory with experimentally observable phenomena. 
Over the past decade, substantial effort has been devoted to describing driven open quantum systems using approaches such as Floquet master equations~\cite{dittrich_driven_1993, breuer_quasistationary_2000} and nonequilibrium Green's function (NEGF) techniques~\cite{stafford_resonant_1996, camalet_current_2003, arrachea_green-function_2005, liu_keldysh_2017}. 
These methods have provided important insights into transport and thermalization in driven systems. 

Despite this progress, the statistical description of periodically driven systems remains less transparent than their equilibrium counterparts. 
In thermal equilibrium, the Fermi-Dirac distribution provides a universal characterization of state occupations. 
For Floquet systems, however, quasienergies are defined only modulo integer multiples of the driving frequency, and the presence of infinitely many Floquet sidebands complicates the notion of a thermal distribution. 
Previous studies have examined Floquet occupations and steady states under various assumptions, such as mutually commuting Hamiltonians~\cite{shirai_condition_2015, matsyshyn_fermi-dirac_2023, peralta_gavensky_stifmmode_2025} or high-frequency modulation~\cite{shirai_effective_2016, liu_keldysh_2017}.
These limitations motivate the search for a rigorous and unified derivation of the statistical distribution for periodically driven systems coupled to a bath.

In this work, we apply the NEGF method to periodically driven quantum systems coupled to baths and derive an analytical expression for the Floquet steady-state distribution.
By combining Floquet theory with the Floquet representation, we derive explicit expressions for the retarded, advanced, and lesser Green's functions and obtain the reduced density matrix governing steady-state populations. 
The resulting distribution takes the form of a weighted sum of Fermi functions shifted by integer multiples of the driving frequency. 
We show that the weights are determined directly by the Fourier components of the micromotion operator, thereby providing a clear physical interpretation of Floquet sideband occupations.
The formulation is not restricted to Hamiltonians that mutually commute at different times.
Furthermore, we demonstrate that the ideal featureless-bath assumption can be relaxed without modifying the resulting Floquet distribution in the weak-coupling limit. 
As an application, we derive a Floquet generalization of the Landauer formula for the DC component of the current in a two‑bath setup, where the equilibrium Fermi functions are replaced by their Floquet‑modified counterparts.
Together, these results establish a coherent description of Floquet quantum statistics and clarify how periodic driving reshapes occupation distributions in open quantum systems.

The remainder of this paper is organized as follows. 
Section II reviews Floquet theory and the Floquet representation used throughout the work. 
Section III introduces the system-bath formalism and derives the relevant Green's functions and self-energies. 
Section IV considers weak system-bath coupling and the wide-band limit, while Section V evaluates the Keldysh equation and derives the Floquet distribution. 
Section VI applies the formalism to transport and develops a Floquet version of the Landauer formula. 
Finally, Section VII summarizes the results and discusses future directions.
\section{Floquet theory and Floquet representation} \label{sec:floquet_thm}

We begin with a brief overview of Floquet theory and the Floquet/Sambe representation, which will be used extensively throughout the rest of the work. 
Floquet's theorem is analogous to Bloch's theorem for a space-periodic lattice potential, where the wavefunction can be expressed as a product of a periodic function and a complex exponential. 
For a Hamiltonian periodic in time, $H(t+2\pi/\Omega) = H(t)$, the evolution operator satisfies the time-dependent Schr\"odinger equation, $i\hbar \partial_t U(t) = H(t) U(t)$.
Floquet's theorem~\cite{bukov_universal_2015, bandyopadhyay_floquet_2022} states that $U(t)$ can be decomposed as, 
\begin{equation}
    U(t) = P(t) \exp\left( \frac{t}{i\hbar} H_F \right) , \label{eq:floquet_thm}
\end{equation}
where the micromotion operator $P(t)$ is unitary and periodic with the same period of $2\pi/\Omega$, and $H_F$ is a time-independent Hermitian quantity called the stroboscopic or Floquet Hamiltonian. 
Since $P(t)$ and $H_F$ are not uniquely defined, we refer to this ambiguity as the Floquet gauge. 
Unfortunately, finding $P$ or $H_F$ is as difficult as solving the time-dependent Schr\"odinger equation in general, and one usually resorts to numerical methods or approximation schemes such as the Magnus expansion~\cite{blanes_magnus_2009, bukov_universal_2015}. 
We bypass this hurdle by adopting the perspective of a Floquet engineer, treating $P(t)$ and $H_F$ as the primary inputs from which $H(t)$ is constructed. 
The initial condition $U(0) = I$ requires $P(0) = I$. 
If the initial time is not $t=0$, the two-time evolution operator 
\begin{align}
    U(t, t') &= U(t) U(t')^\dagger \notag \\
    &= P(t) P(t')^\dagger e^{\frac{t-t'}{i\hbar} P(t') H_F P(t')^\dagger} \label{eq:floquet_thm_general}
\end{align}
should be used instead. 
The form of Eq.~\eqref{eq:floquet_thm_general} mimics Eq.~\eqref{eq:floquet_thm}, implying that $P(t') H_F P(t')^\dagger$ is the Floquet Hamiltonian corresponding to the starting time $t=t'$ instead of $t=0$. 
Thus, the eigenvalues of $H_F$ (quasienergies) are deemed to be more physically meaningful than their eigenstates, as they remain invariant under shifts of the time origin. 
It should be noted that the quasienergies are not unique and can be redefined modulo integer multiples of $\hbar\Omega$. 

When $H(t)$ is time-independent, $U(t, t') = U(t - t')$ is translationally invariant because it depends only on the time difference, and the standard techniques of the Fourier transform apply.
For the present case of a temporally periodic Hamiltonian, Floquet's theorem, Eq.~\eqref{eq:floquet_thm}, yields a weaker condition  
\begin{equation}
    U(t + 2\pi/\Omega, t' + 2\pi/\Omega) = U(t, t') .
\end{equation}
For such two-argument functions with discrete translational symmetry, the Floquet representation~\cite{tsuji_correlated_2008} arises as a natural extension of the Fourier transform. 
The Floquet representation is defined via a discrete Fourier transform in the average variable $(t + t')/2$, and a continuous Fourier transform in the relative variable $t - t'$, namely, 
\begin{align}
    &\quad U(t,t') \to U_{mn}(\omega) \notag \\
    &= \frac{\Omega}{2\pi} \int_0^{\frac{2\pi}{\Omega}} \d t \int_{-\infty}^\infty \d t' \,
    e^{i (m\Omega + \omega) t} e^{-i (n\Omega + \omega) t'} U(t, t') , \label{eq:floquet_rep}
\end{align}
where $m, n \in \mathbb{Z}$. 
It is convenient to regard $U_{mn}(\omega)$ as the $(m,n)$-th block of an infinite matrix, denoted in bold uppercase as $\bm{U}(\omega)$. 
The Floquet representation may be Fourier-inverted by 
\begin{equation}
    U(t, t') 
    = \sum_{mn} \int_{-\frac{\Omega}{2}}^{\frac{\Omega}{2}} \frac{\d \omega}{2\pi} \,
    e^{-i(m\Omega+\omega) t} e^{i(n\Omega+\omega) t'} U_{mn}(\omega) . \label{eq:inverse_floquet}
\end{equation}

Perhaps the most significant advantage of the Floquet representation is the Floquet convolution theorem~\cite{tsuji_correlated_2008,tang_modulating_2024}, which reduces time convolutions into matrix multiplications. 
Given two functions $A$ and $B$ eligible for the Floquet representation, their convolution in Floquet space is
\begin{equation}
    \int_{-\infty}^{\infty} \d t_1 \, A(t, t_1) B(t_1, t') \to \sum_{k} A_{mk}(\omega) B_{kn}(\omega) \text{.} 
\end{equation} 
It is worth noting that if $A$ and $B$ are diagonal matrices that satisfy the usual translational invariance, their Floquet representations are also diagonal and thus commute. 

The Floquet convolution theorem consequently motivates the definition of the Floquet representation for periodic single-time functions. 
Consider casting the following multiplication as a convolution, $P(t) U(t, t') = \int_{-\infty}^{\infty} \d t_1 \, P(t) \delta(t, t_1) U(t_1, t')$. 
It suggests that $P(t)$ can be promoted to a two-time function by $P(t) \delta(t-t')$, and its Floquet representation can accordingly be defined via Eq.~\eqref{eq:floquet_rep}. 
The Floquet representation of single-time functions will not depend on $\omega$, and thus they are just denoted as $\bm{P}$ without any argument. 
Its elements are $P_{mn} = P_{m-n}$, where
\begin{equation}
    P_\xi = \frac{\Omega}{2\pi} \int_0^{\frac{2\pi}{\Omega}} \d t \, e^{i \xi \Omega t} P(t) 
\end{equation}
is the $\xi$-th discrete Fourier component of $P(t)$. 
It is known that periodic driving causes functions of energy or frequency to replicate at shifted arguments, $\omega \to \omega + \xi \Omega$.
We shall later see that the (block) off-diagonal terms of $\bm{P}$ in Eq.~\eqref{eq:floquet_thm} are shown to be directly responsible for the strength of these Floquet shifts in the Floquet distribution.

The last key benefit of the Floquet representation is its treatment of time derivatives. 
Similar to the ordinary Fourier transform, differentiation in the time domain corresponds to multiplication by a certain block-diagonal matrix, whose entries are $\Omega_{mn}(\omega) = (\omega + m \Omega) \delta_{mn} I$. 
Depending on whether the derivative is taken with respect to the first or second argument, $\bm{\Omega}$ is multiplied on the left or right, respectively, as 
\begin{subequations}
\begin{align}
    i \partial_{t} U(t, t') &\to \bm{\Omega}(\omega)\bm{U}(\omega) , \\
    i \partial_{t'} U(t, t') &\to - \bm{U}(\omega) \bm{\Omega}(\omega) .
\end{align}
\end{subequations}
The derivative of a single-time function is a special case and can be carefully worked out (using the product rule) to be the commutator, 
\begin{equation}
    i [\partial_t P(t)] \delta(t-t') \to [\bm{\Omega}(\omega), \bm{P}] .
\end{equation}

An important and relevant application of the Floquet representation is that it enables the numerical evaluation of otherwise intractable quantities by truncating the infinite matrix in the Floquet space~\cite{shirley_solution_1965,rudner_floquet_2020}. 
For example, the Schr\"odinger equation for $P(t)$ defined in Eq.~\eqref{eq:floquet_thm},
\begin{equation}
    H(t) = P(t) H_F P(t)^\dagger + i\hbar \frac{\d P(t)}{\d t} P(t)^\dagger ,
\end{equation}
can be written as a unitary similarity transformation in the Floquet representation, 
\begin{equation}
    \bm{H} - \hbar \bm{\Omega}(\omega) = \bm{P} [\bm{H}_F - \hbar \bm{\Omega}(\omega)] \bm{P}^\dagger . \label{eq:shirley_diagonalization}
\end{equation}
The superscript $\dagger$ indicates conjugate transpose, where the transpose action involves both the original internal matrix indices and the Floquet representation indices. 
Henceforth, we will mostly omit the explicit $\omega$ dependence in the Floquet representation, with the understanding that only functions with a single time argument are truly independent of $\omega$. 
\section{System and bath}

We introduce the various electron Green’s functions starting from the basic principles of quantum mechanics. 
This section serves both to establish the notation used and to provide a systematic recipe for computing the relevant quantities.
We mainly follow the approach outlined by Matsyshyn et al.~\cite{matsyshyn_fermi-dirac_2023}, retaining most of their notation.
We recast the derivation in the NEGF framework and extend the formalism to the general case of a time‑dependent, non‑commuting Hamiltonian with generic system–bath coupling. 

We assume that the quantum system and bath are modeled as a direct sum within the full Hilbert space, 
\begin{equation}
    H(t) = 
    \begin{pmatrix}
        H_S(t) & H_{SB}(t) \\
        H_{BS}(t) & H_B 
    \end{pmatrix} 
    , 
    \ket{\psi} = 
    \begin{pmatrix}
        \ket{\psi_S(t)} \\
        \ket{\psi_B(t)} 
    \end{pmatrix} . \label{eq:schrodinger}
\end{equation} 
The subscript $S$ denotes the system, while the subscript $B$ denotes the bath.
Such a decomposition as a direct sum is meaningful when the system and bath are non-overlapping regions~\cite{matsyshyn_fermi-dirac_2023, campaioli_quantum_2024}. 
A single electron can be either in the system, in the bath, or in a superposition state. 
By contrast, in the open quantum systems community, the system and bath are typically represented as a tensor product of Hilbert spaces, an approach that has also been extended in studies of Floquet dynamics~\cite{kohler_floquet-markovian_1997, hone_statistical_2009, ketzmerick_statistical_2010, iadecola_floquet_2015, shirai_condition_2015, shirai_effective_2016}.
Here, we only consider a non-interacting problem, where the first quantization notation is sufficient. 
Suppose that the evolution operators for the system and bath have been found, which individually satisfy
\begin{subequations}
\begin{align}
    i \hbar \partial_t U_S(t, t') = H_S(t) U_S(t, t') , \\
    i \hbar \partial_t U_B(t-t') = H_B U_B(t-t') .
\end{align}
\end{subequations}
The bath state vector is implicitly given by 
\begin{align}
    \ket{\psi_B(t)} &= U_B(t - t_0) \ket{\psi_B(t_0)} \notag \\
    &\quad+ \frac{1}{i\hbar} \int_{t_0}^t \d t_1 U_B(t-t_1) H_{BS}(t_1) \ket{\psi_S(t_1)} , \label{eq:schrodinger_bath}
\end{align}
where $t_0$ denotes the time at which the initial condition of the quantum state is known. 
Eliminating the bath state from Eq.~\eqref{eq:schrodinger}, the evolution of the system follows a modified Schr\"odinger equation, 
\begin{align}
    i\hbar \partial_t \ket{\psi_S(t)} 
    &= H_S(t) \ket{\psi_S(t)} \notag \\
    &\quad + H_{SB}(t) U_B(t-t_0) \ket{\psi_B(t_0)} \notag \\
    + H_{SB}(t) \frac{1}{i\hbar} & \int_{t_0}^t \d t_1 \, U_B(t-t_1) H_{BS}(t_1) \ket{\psi_S(t_1)} . \label{eq:schrodinger_modified}
\end{align}

Here, it is appropriate to define the retarded self-energy as 
\begin{equation}
    \Sigma^R(t,t') = H_{SB}(t) g^R_B(t-t') H_{BS}(t') , \label{eq:sigmar}
\end{equation}
where 
\begin{equation}
    g^R_B(t) = \frac{1}{i\hbar} \Theta(t) U_B(t) \label{eq:gbr_free}
\end{equation}
is the free Green's function of the bath, and $\Theta$ is the Heaviside step function. 
The advanced counterpart is
\begin{equation}
    \Sigma^A(t, t') = H_{SB}(t) g^A_B(t-t') H_{BS}(t') , \label{eq:sigmaa}
\end{equation}
with
\begin{equation}
    g^A_B(t) = -\frac{1}{i\hbar} \Theta(-t) U_B(t) .
\end{equation}
While the retarded and advanced Green's functions are response functions, there is also a lesser version which governs correlations.
The lesser self-energy $\Sigma^<$ has the same form as Eqs.~\eqref{eq:sigmar} and \eqref{eq:sigmaa}, but will be presented here in the Floquet representation instead, 
\begin{align}
    \bm{\Sigma}^< = \bm{H}_{SB} \bm{g}^<_B \bm{H}_{BS} . \label{eq:sigmaless}
\end{align}
To use the Floquet representation, the coupling Hamiltonian is required to be periodic with the same period as the system Hamiltonian. 
Although this requirement could motivate future investigations, there is no a priori reason to impose it here; therefore, we restrict our attention to static coupling. 

The initially isolated bath is assumed to be in thermal equilibrium at temperature $T$, so $g^<_B$ can be determined by the fluctuation-dissipation theorem, which is expressed in the frequency domain as 
\begin{equation}
    g^<_B(\omega) = - f(\hbar\omega) [g^R_B(\omega) - g^A_B(\omega)] . \label{eq:bath_fdt}
\end{equation}
$f(E) = [1 + \exp(\beta (E-\mu))]^{-1}$ is the Fermi-Dirac function with inverse temperature $\beta = 1/k_B T$ and chemical potential $\mu$. 
Since 
\begin{equation}
    g^R_B(\omega) - g^A_B(\omega) = \frac{1}{i\hbar} U_B(\omega)
\end{equation}
is essentially a collection of Dirac-Delta functions, Eq.~\eqref{eq:bath_fdt} can alternatively be written as 
\begin{equation}
    g^<_B(\omega) = - \rho_{B}(t_0) [g^R_B(\omega) - g^A_B(\omega)] ,
\end{equation}
where $\rho_B(t_0) = f(H_B)$ means that the bath is assumed to be in thermal equilibrium initially at $t_0$. 
Strictly speaking, $\rho_B(t_0)$ is not a density operator since its trace is not unity. 
Fermi-Dirac statistics naturally emerge from employing second quantization with fermionic operators. 
Indeed, $\rho_B(t_0)$ can be interpreted as the projection of the density operator in Fock space onto the one-particle bath states. 
In the time domain, $g^<_B$ is thus 
\begin{equation}
    g^<_B(t) = -\frac{1}{i\hbar} \rho_B(t_0) U_B(t) .
\end{equation}

Returning to the evolution of the system, Eq.~\eqref{eq:schrodinger_modified} can now be recast as
\begin{align}
    &\quad [i\hbar \partial_t - H_S(t)] \ket{\psi_S(t)} 
    - \int_{t_0}^\infty \d t_1 \, \Sigma^R(t,t_1) \ket{\psi_S(t_1)} \notag \\
    &= H_{SB}(t) U_B(t - t_0) \ket{\psi_B(t_0)} . \label{eq:schrodinger_selfenergy}
\end{align}
We can extend the upper limit of the integral to $\infty$ due to the step function in $\Sigma^R$. 
The left-hand side of Eq.~\eqref{eq:schrodinger_selfenergy} is a linear differential operator, and its Green's function~\cite{pan_asymmetry-induced_2025} satisfies the Dyson equation 
\begin{align}
    &\quad [i\hbar \partial_t - H_S(t)] G_S^R(t, t') 
    - \int_{-\infty}^{\infty} \d t_1 \, \Sigma^R(t, t_1) G_S^R(t_1, t') \notag \\
    &= \delta(t-t') I . 
\end{align}
There is a corresponding advanced version, $G_S^A$, which will be used later and defined similarly as follows,
\begin{align}
   &\quad [i\hbar \partial_t - H_S(t)] G_S^A(t, t') 
    -  \int_{-\infty}^{\infty} \d t_1 \, \Sigma^A(t, t_1) G_S^A(t_1, t') \notag \\
   &= \delta(t-t') I . 
\end{align}
One method to solve for the Green's functions is via their Floquet representations,
\begin{subequations}
\begin{align}
    \bm{G}_S^R &= (\hbar \bm{\Omega} - \bm{\Sigma}^R - \bm{H}_S)^{-1} , \label{eq:gsr_floquet} \\
    \bm{G}_S^A &= (\hbar \bm{\Omega} - \bm{\Sigma}^A - \bm{H}_S)^{-1} .
\end{align}
\end{subequations}
Physically, Eq.~\eqref{eq:gsr_floquet} represents a Dyson equation, while mathematically it corresponds to the inverse of a Schur complement~\cite{thakur_tutorial_2023}, 
\begin{equation}
    \begin{pmatrix}
        \bm{G}_S^R & \bm{G}_{SB}^R \\
        \bm{G}_{BS}^R & \bm{G}_B^R 
    \end{pmatrix}
    = 
    \begin{pmatrix}
        \hbar \bm{\Omega} + i\eta - \bm{H}_S & - \bm{H}_{SB} \\
        - \bm{H}_{BS} & \hbar \bm{\Omega} + i\eta - \bm{H}_B
    \end{pmatrix}^{-1} .
\end{equation} 
The infinitesimal positive constant $\eta$ is a standard mathematical trick to preserve causality~\cite{stefanucci_nonequilibrium_2013}, as otherwise the inverse operation is ill-defined for a singular matrix. 
Since we have already defined the bath's free Green's function as $\bm{g}_B^R(\omega) = (\hbar \bm{\Omega} + i\eta - \bm{H}_B)^{-1}$ in Eq.~\eqref{eq:gbr_free}, the system's free Green's function can likewise be defined as $\bm{g}_S^R(\omega) = (\hbar \bm{\Omega} + i\eta - \bm{H}_S)^{-1}$. 
The full Green's functions can then be solved exactly as 
\begin{subequations}
\begin{align}
    \bm{G}_S^R &= ((\bm{g}_S^R)^{-1} - \bm{\Sigma^R})^{-1} , \\
    \bm{G}_B^R &= \bm{g}_B^R + \bm{g}_B^R \bm{H}_{BS} \bm{G}_S^R \bm{H}_{SB} \bm{g}_B^R , \label{eq:gbr_floquet} \\
    \bm{G}_{SB}^R &= \bm{G}_S^R \bm{H}_{SB} \bm{g}_B^R , \\
    \bm{G}_{BS}^R &= \bm{g}_B^R \bm{H}_{BS} \bm{G}_S^R .
\end{align}
\end{subequations}

Returning to the solution of Eq.~\eqref{eq:schrodinger_selfenergy}, taking the convolution with the Green's function in the time domain yields    
\begin{equation}
    \ket{\psi_S(t)} = \int_{-\infty}^{\infty} \d t_1 \,
    G_S^R(t, t_1) H_{SB}(t_1) U_B(t_1 - t_0) \ket{\psi_B(t_0)} . \label{eq:system_particular}
\end{equation}
If one is interested in the evolution of the bath state $\ket{\psi_B(t)}$, a similar result can be obtained by substituting Eq.~\eqref{eq:system_particular} into Eq.~\eqref{eq:schrodinger_bath},
\begin{align}
    \ket{\psi_B(t)} &= U_B(t-t_0) \ket{\psi_B(t_0)} \notag \\
    &\quad+ \int_{t_0}^{\infty} \d t_1 \int_{-\infty}^{\infty} \d t_2 \,
    g^R_B(t-t_1) H_{BS}(t_1) \notag \\
    &\quad\times G_S^R(t_1, t_2) H_{SB}(t_2) U_B(t_2-t_0) \ket{\psi_B(t_0)} . \label{eq:bath_particular}
\end{align}

The density matrix formalism is useful for describing mixed states such as those arising from a thermal ensemble. 
Using two copies of Eq.~\eqref{eq:system_particular}, the density matrix for a pure state projected onto the system states is
\begin{align}
    \ket{\psi_S(t)} \bra{\psi_S(t)} = \int_{-\infty}^{\infty} \d t_1 \,
    G_S^R(t, t_1) H_{SB}(t_1) U_B(t_1 - t_0) \notag \\
    \times \ket{\psi_B(t_0)} \bra{\psi_B(t_0)} 
    \int_{-\infty}^{\infty} \d t_2 \,
    U_B(t_0 - t_2) H_{BS}(t_2) G_S^A(t_2, t) .
\end{align}
The general result follows by taking linear combinations, 
\begin{align}
    \rho_S(t) = \int_{-\infty}^{\infty} \d t_1 \,
    G_S^R(t, t_1) H_{SB}(t_1) U_B(t_1 - t_0) \notag \\
    \times \rho_B(t_0) \int_{-\infty}^{\infty} \d t_2 \,
    U_B(t_0 - t_2) H_{BS}(t_2) G_S^A(t_2, t) . \label{eq:density_matrix_s}
\end{align}
Since $\rho_B(t_0)$ is assumed to commute with $U_B$, we identify 
\begin{equation}
    U_B(t_1 - t_0) \rho_B(t_0) U_B(t_0 - t_2) = -i \hbar g^<_B(t_1 - t_2) , \label{eq:gless_idenitfy}
\end{equation}
which erases the dependence on $t_0$. 
Using Eqs.~\eqref{eq:sigmaless} and \eqref{eq:gless_idenitfy}, Eq.~\eqref{eq:density_matrix_s} can be simplified as
\begin{equation}
    \rho_S(t) = -i\hbar \iint_{-\infty}^{\infty} \d t_1 \d t_2 \,
    G_S^R(t, t_1) \Sigma^<(t_1, t_2) G_S^A(t_2, t) . 
\end{equation}
Thus, the reduced density matrix, which governs the population statistics, can be obtained from the lesser Green's function by
\begin{equation}
    \rho_S(t) = -i \hbar G_S^<(t, t) . \label{eq:population}
\end{equation}
The lesser Green's function is, in turn, calculated using the Keldysh equation~\cite{haug_quantum_2008},
\begin{equation}
    \bm{G}_S^< = \bm{G}_S^R \bm{\Sigma}^< \bm{G}_S^A . \label{eq:keldysh}
\end{equation}
Equation~\eqref{eq:keldysh} is presented in the Floquet representation, since we shall subsequently understand that its evaluation is more convenient there than in the time domain.
To return to the time domain as required in Eq.~\eqref{eq:population}, Eq.~\eqref{eq:inverse_floquet} can be combined with the Floquet modular property, $G_{mn}(\omega+k\Omega) = G_{(m+k)(n+k)}(\omega)$, to yield 
\begin{equation}
    G(t, t) 
    = \sum_{m} e^{-i m \Omega t} \int_{-\infty}^{\infty} \frac{\d \omega}{2\pi} \, G_{m0}(\omega) . \label{eq:back_equal_time}
\end{equation}
This is a suitable simplification for analytical calculations, but it poses challenges for numerical computations, as the integral must account for multiple poles.

The two-point electron Green's functions are usually presented using the second-quantized notation~\cite{haug_quantum_2008,stefanucci_nonequilibrium_2013,wang_transport_2023}, 
\begin{subequations}
\begin{align}
    G_{S,ij}^R(t, t') &= \frac{1}{i\hbar} \Theta(t-t') \langle \{c_i(t), c_j(t')^\dagger\} \rangle , \\
    G_{S,ij}^A(t, t') &= -\frac{1}{i\hbar} \Theta(t'-t) \langle \{c_i(t), c_j(t')^\dagger\} \rangle , \\
    G_{S,ij}^<(t, t') &= - \frac{1}{i\hbar} \langle c_j(t')^\dagger c_i(t) \rangle ,
\end{align}
\end{subequations}
where $\langle \ldots \rangle$ is the ensemble average, $\{ \ldots \}$ is the anti-commutator, $c_i(t)$ and $c_j(t')^\dagger$ are the second-quantized fermionic operators in the Heisenberg picture. 
In the general case, the self-energy is expressed in terms of higher-order Green's functions, and this creates a hierarchical structure that continues indefinitely. 
For the non-interacting case studied here, the Dyson equation for the two-point Green's function is fortunately closed, and it can be verified that the various definitions are consistent. 

We conclude this section by briefly discussing the other sectors of the full lesser Green's function. 
The expression for $\bm{G}_B^R$ is already presented in Eq.~\eqref{eq:gbr_floquet}.
To obtain the lesser version, the density matrix arising from the bra and ket version of Eq.~\eqref{eq:bath_particular} yields four terms, which consolidate to 
\begin{align}
    \bm{G}^<_B = \bm{g}^<_B 
    + \bm{g}^R_B \bm{H}_{BS} \bm{G}_S^R \bm{H}_{SB} \bm{g}^<_B \notag \\
    + \bm{g}^R_B \bm{H}_{BS} \bm{G}_S^< \bm{H}_{SB} \bm{g}^A_B \notag \\ 
    + \bm{g}^<_B \bm{H}_{BS} \bm{G}_S^A \bm{H}_{SB} \bm{g}^A_B .
\end{align}
Evidently, the results can be readily obtained with Langreth's rule~\cite{haug_quantum_2008,stefanucci_nonequilibrium_2013,wang_transport_2023}. 
Likewise, the lesser components of the mixed Green's functions are   
\begin{subequations}
\begin{align}
    \bm{G}_{SB}^< 
    &= \bm{G}_S^R \bm{H}_{SB} \bm{g}_B^<
    + \bm{G}_S^< \bm{H}_{SB} \bm{g}_B^A , \\
    \bm{G}_{BS}^< 
    &= \bm{g}_B^R \bm{H}_{BS} \bm{G}_S^< 
    + \bm{g}_B^< \bm{H}_{BS} \bm{G}_S^A .
\end{align}
\end{subequations}
\section{Weakly coupled bath}

\begin{figure}[htb]
    \centering
    \includegraphics[width=\linewidth]{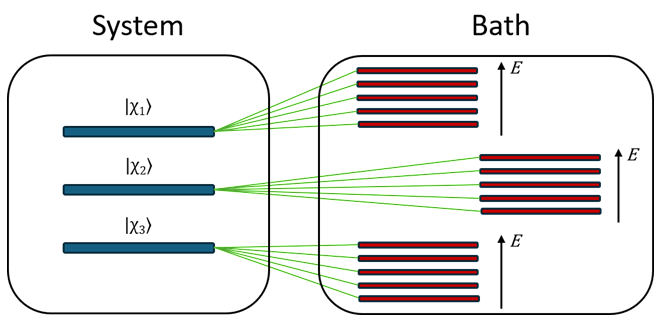}
    \caption{
    \label{fig:connection} 
    Schematic of the system-bath coupling for a hypothetical system with $N_S = 3$ basis states and a bath with $15$ states. 
    The bath depicted is composed of three identical sub‑baths, with each containing five levels. 
    Each system level couples exclusively and fully to one sub-bath. 
    In the ideal bath assumption, both the coupling strength and the sub-bath density of states are uniform. 
    }
\end{figure}

We proceed by considering a specific class of system-bath coupling to obtain concrete results. 
The time-dependent system Hamiltonian, in its most general form, can be written as
\begin{equation}
    H_S(t) = \sum_{mn} \epsilon_{mn}(t) \ket{\chi_m} \bra{\chi_n} .
\end{equation}
This goes beyond the diagonal case considered in Ref.~\cite{matsyshyn_fermi-dirac_2023}, which implies that the Hamiltonians at any two times commute. 
Without loss of generality, we work in the eigenbasis of $H_F$, which means that the chosen orthonormal states satisfy $H_F \ket{\chi_j} = [H_F]_{jj} \ket{\chi_j}$. 
The system states are consistently labeled by $\chi$, while the bath states will be denoted with $\varphi$. 
Next, the bath Hamiltonian is written in a diagonal bath basis, 
\begin{equation}
    H_B = \sum_{m\alpha} \varepsilon_{\alpha} \ket{\varphi_{m \alpha}} \bra{\varphi_{m \alpha}} . \label{eq:general_hb}
\end{equation}
When expressing the Hamiltonians as a sum over two indices, we label the basis of system states with Latin indices, and reserve the Greek indices for the bath states. 
Thus, the discrete index $m$ in Eq.~\eqref{eq:general_hb} takes $N_S$ values, where $N_S$ is the dimension of $H_S$. 
We stick to the notation used in Ref.~\cite{matsyshyn_fermi-dirac_2023} for ease of cross-reference, where the matrix elements of the system and the bath Hamiltonians are denoted by $\epsilon$ and $\varepsilon$, respectively, despite their visual similarity. 
Essentially, Eq.~\eqref{eq:general_hb} describes a direct sum of $N_S$ copies of the same bath sub-Hamiltonian. 
Lastly, the system and bath are assumed to be coupled by an interaction Hamiltonian, 
\begin{align} 
    H_{SB} = \sum_{m \alpha} \lambda_\alpha \ket{\chi_m} \bra{\varphi_{m \alpha}} ,
\end{align}
and $H_{BS} = H_{SB}^\dagger$. 
Each system state is fully coupled to a separate copy of the bath sub-Hamiltonian. 
The setup is shown schematically in Fig.~\ref{fig:connection}. 

With all the variables introduced, the summation over the bath states can be approximated by an integral with the density of states. 
As noted by Leggett et al.~\cite{leggett_dynamics_1987}, the effect of the bath is usually encapsulated in its density of states or spectral function. 
We shall assume that the bath energy levels are sufficiently dense for the spectral function to be smooth, 
\begin{equation}
    \sum_\alpha |\lambda_\alpha|^2 \delta \left(\omega - \frac{\varepsilon_\alpha}{\hbar} \right) \to \frac{\hbar}{2 \pi} \Gamma(\omega) . \label{eq:bath_spectral}
\end{equation}
The normalization differs from Ref.~\cite{matsyshyn_fermi-dirac_2023} by a factor of one-half, in anticipation of the standard convention $\Gamma = i(\Sigma^R - \Sigma^A)$~\cite{datta_mesoscopic_1995}. 
Essentially, this translates into the following continuum approximation. 
For some test function $h$, 
\begin{equation}
    \sum_{\alpha} |\lambda_\alpha|^2 h(\varepsilon_\alpha) \to \int_{-\infty}^{\infty} \frac{\hbar \d \omega}{2 \pi} \, \Gamma(\omega) h(\hbar\omega) .
\end{equation}
If $\lambda_\alpha$ denotes a weak coupling, then the broadening term $\Gamma(\omega) \sim \lambda_\alpha^2$ is also a small quantity. 

The retarded self-energy can then be computed using Eq.~\eqref{eq:sigmar} as 
\begin{align}
    \Sigma^R(t, t') 
    &= \sum_{\alpha} |\lambda_\alpha|^2 \frac{1}{i\hbar} \Theta(t-t') e^{\frac{t-t'}{i\hbar} \varepsilon_\alpha} \notag \\
    &\to - i \Theta(t-t') \Gamma(t-t') . \label{eq:general_sigmar}
\end{align}
We adopt the convention in which a function and its Fourier transform are denoted by the same symbol, with the distinction made through the argument. 
Note that although $\Sigma^R$ has dimensions of $N_S \times N_S$, we omit writing the identity matrix factor in the interest of brevity. 
Similarly, the advanced version is 
\begin{equation}
    \Sigma^A(t, t') = i \Theta(t'-t) \Gamma(t-t') .
\end{equation}

To proceed, a particular form for $\Gamma(\omega)$ should be chosen. 
The wide-band approximation is commonly employed in models of dissipation and decoherence~\cite{raju_quantum_2016, haughian_quantum_2018, matsyshyn_fermi-dirac_2023}, where the bath's density of states is constant and spans all energies. 
In Eq.~\eqref{eq:bath_spectral}, we can therefore assume that the coupling is independent of the index $\alpha$, leading to a broad and uniform spectral function $\Gamma(\omega) = \Gamma$. 
In the time domain, this is $\Gamma(t) = \delta(t) \Gamma$. 
From the symmetry consideration $\Theta(-t) + \Theta(t) = 1$, the step function is required to take the value of $1/2$ when its argument is $0$. 
Thus, the self-energies for the infinite featureless bath can be summarized as 
\begin{subequations}
\begin{align}
    \Sigma^R(t, t') &= -i \frac{\Gamma}{2} \delta(t-t') , \quad 
    \Sigma^R(\omega) = -i \frac{\Gamma}{2} ,\label{eq:sigmar_featureless} \\
    \Sigma^A(t, t') &= i \frac{\Gamma}{2} \delta(t-t') , \quad
    \Sigma^A(\omega) = i \frac{\Gamma}{2} .
\end{align}
\end{subequations}
To be more rigorous, one can consider the family of Lorentzian functions $\Gamma(\omega) = \Gamma/(1 + \omega^2 \tau^2)$, parameterized by the Drude timescale $\tau$.
The corresponding inverse Fourier transform is $\Gamma(t) = (\Gamma/2\tau) \exp(-|t|/\tau)$. 
The self-energy takes an analytical form, $\Sigma^R(\omega) = \Gamma/2(\omega \tau + i)$, and the Kramers-Kronig relation is duly satisfied. 
We take the limit $\tau \to  0^+$ to recover Eq.~\eqref{eq:sigmar_featureless}. 

\section{Evaluating the Keldysh equation}

In the Floquet representation, the self-energies of the ideal bath are scalar multiples of the identity and will thus commute with all other quantities. 
Only with the special choice of self-energy proportional to the identity can Eq.~\eqref{eq:gsr_floquet} be simultaneously diagonalized (in a manner similar to Eq.~\eqref{eq:shirley_diagonalization}), leading to the following, 
\begin{subequations}
\begin{align}
    \bm{P}^\dagger \bm{G}_S^R \bm{P} &= (\hbar \bm{\Omega} + i \Gamma/2 - \bm{H}_F)^{-1} , \label{eq:gr_general} \\
    \bm{P}^\dagger \bm{G}_S^A \bm{P} &= (\hbar \bm{\Omega} - i \Gamma/2 - \bm{H}_F)^{-1} . \label{eq:ga_general} 
\end{align}
\end{subequations}
Compared to Eq.~\eqref{eq:gsr_floquet}, the advantage of this form lies in the ease of taking the inverse, since the right-hand side is diagonal ($H_F$ is diagonal by assumption). 

To evaluate the population using $\bm{G}_S^<$, $\bm{\Sigma}^<$ is also needed in the Keldysh equation, Eq.~\eqref{eq:keldysh}. 
Using Eq.~\eqref{eq:sigmaless}, $\bm{\Sigma}^<$ can be cast in a form reminiscent of the fluctuation–dissipation theorem, 
\begin{equation}
    \bm{\Sigma}^< 
    = - \bm{F} [\bm{\Sigma}^R - \bm{\Sigma}^A] 
    = i \Gamma \bm{F}, \label{eq:sigmaless_general}
\end{equation}
where $\bm{F}(\omega)$ is the Floquet representation of the Fermi-Dirac function, with elements 
\begin{equation}
    F_{mn}(\omega) = f(\hbar\omega + m\hbar\Omega) \delta_{mn} I .
\end{equation}
The lesser Green's function is obtained from the Keldysh equation, Eq.~\eqref{eq:keldysh}, and we label components by the Floquet indices $(m,n \in \mathbb{Z})$ as well as the internal matrix indices $(a, b \in \{1, \ldots, N_S\})$, 
\begin{align}
    [(\bm{P}^\dagger \bm{G}_S^< \bm{P})_{mn}]_{ab} 
    &= i \Gamma 
    (\hbar\omega + m \hbar\Omega + i\Gamma/2 - [H_F]_{aa})^{-1} \notag \\
    &\quad\times \sum_{l} f(\hbar\omega + l \hbar \Omega) [P_{l-m}^\dagger P_{l-n}]_{ab} \notag \\
    &\quad\times (\hbar\omega + n\hbar\Omega - i\Gamma/2 - [H_F]_{bb})^{-1} .  \label{eq:gless_featureless_components}
\end{align}
Up to this point, $\Gamma$ has been kept exact. 
Eventually, the weak coupling limit $\Gamma \to 0^+$ is taken, corresponding to approaching $H_{SB} = 0$ along a specific path. 
However, as we will demonstrate, all distinct paths of vanishing coupling lead to the same conclusion. 
The prefactor of $\Gamma$ in Eq.~\eqref{eq:sigmaless_general} appears to suggest that $G_S^<$ vanishes in the infinitesimal $\Gamma$ limit, but that is false in view of the following Sokhotski–Plemelj identity, 
\begin{align}
    &\quad \frac{\eta}{(\omega-\Omega_a+i\eta)(\omega-\Omega_b-i\eta)} \notag \\
    &= \mathcal{O}(\eta) + \begin{cases}
        \pi \delta(\omega - \Omega_a) &\text{if }\Omega_a = \Omega_b , \\
        0 &\text{if }\Omega_a \neq \Omega_b .
    \end{cases} \label{eq:sokhotsky_plemlj_variant}
\end{align} 
For steady-state distribution calculations, only the zeroth‑order result is required, though in other applications the first‑order result may also be needed.

Consider the degenerate $\Omega_a = \Omega_b$ case in Eq.~\eqref{eq:sokhotsky_plemlj_variant}, which implies that the quasienergies differ by integer multiples of the driving frequency, $[H_F]_{aa} - [H_F]_{bb} = k \hbar \Omega$. (This includes the trivial case of $a = b$ and $k = 0$.) 
Thus, to zeroth order in $\Gamma$, 
\begin{align}
    [(\bm{P}^\dagger \bm{G}_S^< \bm{P})_{mn}]_{ab} 
    = \delta_{m-n, k} \frac{2 \pi i}{\hbar}
    \delta(\omega + m \Omega - [H_F]_{aa}/\hbar) \notag \\
    \times \sum_{\xi} f([H_F]_{aa} + \xi \hbar\Omega) [P_\xi^\dagger P_{\xi+k}]_{ab} .\label{eq:gless_featureless_components2}
\end{align}
Here and henceforth, the renamed integer index $\xi = l - m$ runs over all integers. 
Returning to the time domain, Eq.~\eqref{eq:gless_featureless_components2} is 
\begin{align}
    &\quad [P(t)^\dagger G_S^<(t,t) P(t)]_{ab} 
    \notag \\
    &= e^{-i k \Omega t} \frac{-1}{i\hbar} \sum_{\xi} f([H_F]_{aa} + \xi \hbar \Omega) [P_\xi^\dagger P_{\xi+k}]_{ab} . 
\end{align}
Recognizing that $e^{i k \Omega t} = [e^{-\frac{t}{i\hbar} H_F}]_{aa} [e^{\frac{t}{i\hbar} H_F}]_{bb}$, and invoking Floquet's theorem, we obtain 
\begin{align}
    &\quad [U_S(t)^\dagger G_S^<(t,t) U_S(t)]_{ab} 
    \notag \\
    &= \frac{-1}{i\hbar} \sum_{\xi} f([H_F]_{aa} + \xi \hbar \Omega) [P_\xi^\dagger P_{\xi+k}]_{ab} . \label{eq:gless_equal_time}
\end{align}

The density matrix obtained using Eq.~\eqref{eq:population} is thus 
\begin{equation}
    \rho_S(t) = U_S(t) \rho_F U_S(t)^\dagger , \label{eq:rhos}
\end{equation}
where a key result is the analytical expression for the Floquet distribution,
\begin{equation}
    [\rho_F]_{ab} = \sum_{\xi} f([H_F]_{aa} + \xi \hbar \Omega) [P_\xi^\dagger P_{\xi+k}]_{ab} . \label{eq:floquet_distribution}
\end{equation}
Diagonal elements of a density matrix represent occupation, while off-diagonal elements encode coherence. 
If the quasienergies $[H_F]_{aa}$ and $[H_F]_{bb}$ are otherwise nondegenerate, the zeroth-order result will simply be $[\rho_F]_{ab} = 0$. 
If the weak-coupling limit is not used, the resulting time-dependent $\rho_F$ can still be solved exactly (Appendix~\ref{app:lerch_digamma}) as
\begin{align}
    [\rho_F(t)]_{ab} 
    &= \sum_{k \xi} e^{-i (\frac{[H_F]_{bb} - [H_F]_{aa}}{\hbar} + k \Omega) t} \notag \\
    &\quad \times \frac{i\Gamma}{[H_F]_{bb} - [H_F]_{aa} + k\hbar\Omega + i\Gamma}
    [P^\dagger_{\xi} P_{\xi+k}]_{ab} \notag \\
    &\quad \times \frac{1}{2} \biggl[f_{\Gamma/2}([H_F]_{aa} + \xi\hbar\Omega) \notag \\
    &\qquad + f_{\Gamma/2}([H_F]_{bb} + (\xi+k) \hbar\Omega)^* \biggr] . \label{eq:distribution_exact}
\end{align}
Following Matsyshyn et al.~\cite{matsyshyn_fermi-dirac_2023}, we define the auxiliary Fermi function as
\begin{equation}
    f_\eta(\epsilon) = \frac{1}{2} + \frac{i}{\pi} \Psi \left(
        \frac{1}{2} + \beta \frac{\eta + i (\epsilon - \mu)}{2 \pi}
    \right) , \label{eq:fermi_auxiliary}
\end{equation}
with $\Psi$ being the digamma function. 
The equilibrium Fermi function is recovered by $\lim_{\eta \to 0^+} \Re f_\eta(\epsilon) = f(\epsilon)$. 

Leaving out degeneracy simplifies the analysis, as all off‑diagonal terms in $\rho_F$ vanish.
More precisely, energy differences should be significantly greater than $\Gamma$ to avoid near degeneracies~\cite{hone_statistical_2009}. 
We caution that such an assumption would not hold if there are symmetry-protected degeneracies or if the Floquet-Brillouin zone becomes crowded owing to a large number of system states. 
On the other hand, degeneracy can be deliberately introduced, such as by tuning the driving frequency to match an unperturbed energy difference. 

In the nondegenerate case, the above Floquet steady-state distribution reduces to the diagonal Floquet occupation matrix~\cite{matsyshyn_fermi-dirac_2023,shi_floquet_2024,peralta_gavensky_stifmmode_2025, kumari_josephson-current_2024}, 
\begin{equation}
    \rho_F = \sum_{\xi} f(H_F + \xi \hbar \Omega) \mathcal{D}(P_\xi^\dagger P_\xi) . \label{eq:floquet_distribution_diagonal}
\end{equation}
The notation 
\begin{equation}
    \mathcal{D}(P_\xi^\dagger P_\xi) = \sum_j \ket{\chi_j} \bra{\chi_j} P_\xi^\dagger P_\xi \ket{\chi_j} \bra{\chi_j}
\end{equation}
denotes the operation of retaining only the diagonal elements of $P_\xi^\dagger P_\xi$, which also corresponds to the squared norms of the columns of $P_\xi$ (in the eigenbasis of $H_F$). 
The same Floquet distribution can also be derived from the Keldysh equation in the time domain as a further consistency check (Appendix~\ref{app:keldysh_time}). 
The diagonal elements of the Floquet distribution are essentially a weighted sum of horizontally translated Fermi functions, and the Parseval theorem, $\sum_\xi P_\xi^\dagger P_\xi = I$, ensures the normalization of the weights. 

The notion of Floquet sidebands provides an intuitive explanation of the Floquet distribution without invoking any bath. 
The evolution of the Floquet eigenstate $\ket{\chi_j}$ is given by
\begin{equation}
    U_S(t) \ket{\chi_j} = \sum_{\xi} e^{-i \left(\frac{[H_F]_{jj}}{\hbar} + \xi \Omega\right) t} P_{\xi} \ket{\chi_j} .
\end{equation}
Such a linear combination with time-dependent exponential factors resembles a quantum state evolving under a time-independent Hamiltonian, decomposed into stationary states.
Thus, each Floquet replica $P_{\xi} \ket{\chi_j}$ can be identified with energy $[H_F]_{jj} + \xi \hbar\Omega$ and weight $[P_{\xi}^\dagger P_\xi]_{jj}$, which agrees exactly with the Floquet distribution in Eq.~\eqref{eq:floquet_distribution_diagonal}. 
Physically, these weights describe how the periodic micromotion distributes the spectral weight of a Floquet state among different sidebands, so that its steady-state occupation is a weighted average of the reservoir's Fermi functions evaluated at the corresponding sideband energies.
We emphasize that although the quasienergy is not unique, the weight associated with each sideband is unambiguous. 
Consequently, the Floquet distribution is well-defined (such as being independent of the Floquet gauge or the time origin), as required for a physically measurable quantity. 

The analytical Floquet distribution also clarifies the connection with Kohn's periodic thermodynamics~\cite{kohn_periodic_2001}. 
In Kohn's formulation, the steady-state occupations satisfy rate equations whose transition probabilities include contributions from all Floquet sidebands, leading to a generally non-Boltzmann distribution. 
Within the present weak-coupling NEGF framework, these occupations can be analytically evaluated as Eq.~\eqref{eq:floquet_distribution_diagonal}.
When the driving vanishes, $P_\xi \to \delta_{\xi 0} I$, and the equilibrium Fermi distribution is restored.
The micromotion being dominated by a single Fourier component recovers Kohn's approximate result that the occupations reduce to a Boltzmann distribution of the appropriate Floquet branch.
Shirai et al.~\cite{shirai_condition_2015} made a similar approximation, assuming a single term dominates the transition probabilities to satisfy the detailed balance condition.
In comparison, our derivation retains all Floquet sidebands exactly, and without assuming that the Hamiltonian commutes at different times. 

\begin{figure}[htb]
    \centering
    \includegraphics[width=\linewidth]{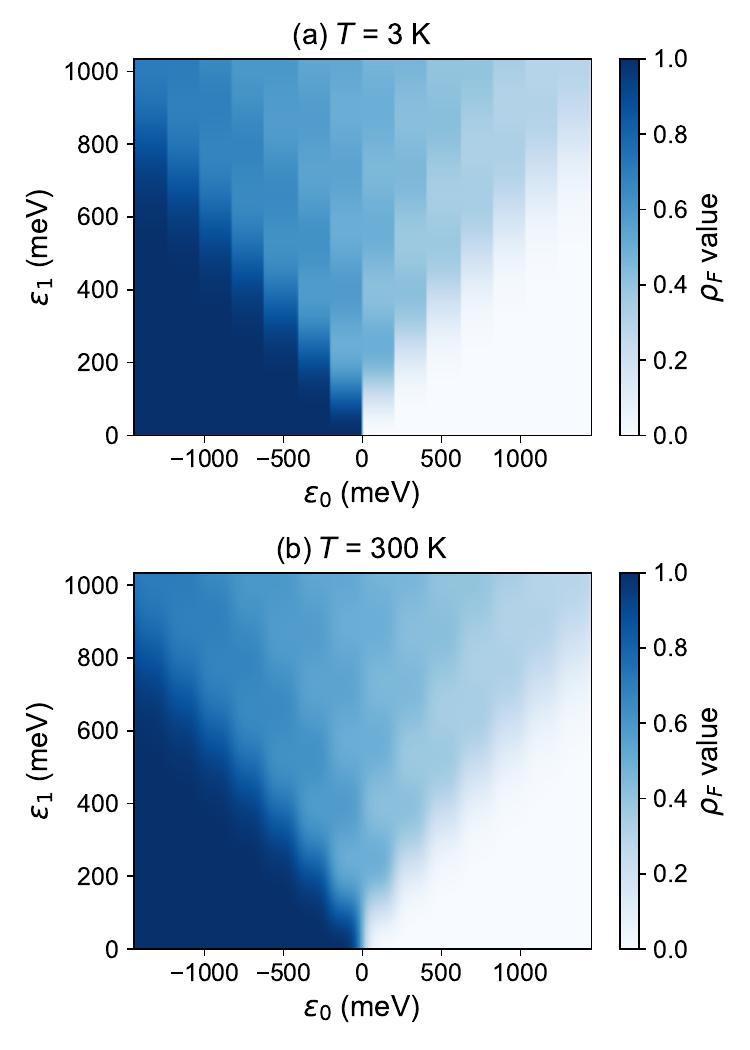}
    \caption{
    \label{fig:floquet_fermi} 
    Heatmaps of the Floquet Fermi function (Eq.~\eqref{eq:floquet_distribution_1d}) at temperatures (a) \SI{3}{\kelvin} and (b) \SI{300}{\kelvin} corresponding to a simple sinusoidal driving (Eq.~\eqref{eq:sinusoidal_hamiltonian}) with frequency $\hbar\Omega = \SI{207}{\milli\electronvolt}$. 
    The horizontal axis denotes the equilibrium energy level, while the vertical axis denotes (half of) the modulation amplitude. 
    When $\varepsilon_1 = 0$, which corresponds to a vanishing driving amplitude, the system is in equilibrium. 
    A stronger drive widens the boundary between populated (low energy) and unpopulated (high energy) regions. 
    At higher temperatures, temperature smearing smooths out the staircase feature of the Floquet Fermi function.
    }
\end{figure}

As an analytic example, we use the simplest periodic Hamiltonian, 
\begin{equation}
    H_S(t) = \epsilon_0 + 2 \epsilon_1 \cos \Omega t , \label{eq:sinusoidal_hamiltonian}
\end{equation}
corresponding to a single resonant level being driven sinusoidally. 
The Jacobi-Anger expansion expresses the Fourier components of $P(t) = \exp(\frac{2 \epsilon_1}{i\hbar \Omega} \sin \Omega t)$ as $P_\xi = J_\xi(\frac{2 \epsilon_1}{\hbar\Omega})$, where $J_\xi$ is the Bessel function of order $\xi$.
The Floquet distribution is then
\begin{equation}
    \rho_S(t)  
    = \sum_{\xi} f(\epsilon_0 + \xi \hbar\Omega) 
    J_{\xi} \left(\frac{2 \epsilon_1}{\hbar\Omega} \right)^2 . \label{eq:floquet_distribution_1d}
\end{equation}
There is no time-dependent rotating basis effect since all $1 \times 1$ matrices commute. 
Equation~\eqref{eq:floquet_distribution_1d} is numerically plotted in Fig.~\ref{fig:floquet_fermi} to illustrate the effect of varying the driving amplitude.
In general, the value of the Floquet Fermi function at a particular energy deviates from its equilibrium counterpart. 
If one attempts to infer the temperature from the resulting state occupancies, this leads to the notion of an effective temperature~\cite{ketzmerick_statistical_2010,pan_asymmetry-induced_2025}. 

Next, we generalize to the case of a weakly coupled, nonfeatureless bath. 
In certain physical scenarios, the wide-band approximation may not accurately describe the system-bath interaction. 
Although assuming a constant density of states and uniform coupling simplifies the analysis, these assumptions can overlook important spectral features of the environment, such as frequency-dependent couplings or cutoffs. 
Specifically, this means that Eq.~\eqref{eq:sigmar_featureless} is not assumed to hold. 
The consequence is that the spectral function is no longer proportional to the identity in the Floquet representation, and the Keldysh equation becomes
\begin{align}
    \bm{G}_S^< = 
    (\hbar \bm{\Omega} - \bm{\Sigma}^R - \bm{H}_S)^{-1} 
    (i \bm{F} \bm{\Gamma})
    (\hbar \bm{\Omega} - \bm{\Sigma}^A - \bm{H}_S)^{-1} . 
\end{align}
Here, $\bm{\Gamma}(\omega)$ is still diagonal in the Floquet representation, and each block on the diagonal is still proportional to the identity, but as a whole, it is not proportional to the Floquet identity.
At first glance, two difficulties emerge.
Firstly, $\bm{\Sigma}^R$ is diagonal in the Floquet representation, but not $\bm{g}_S^R$. 
Thus, there is no common basis to use for taking the inverse. 
Secondly, $\Gamma(\omega)$ now depends on $\omega$, so it is not clear that the Sokhotski-Plemelj formula still applies. 
These concerns can be mitigated by generalizing the Sokhotski-Plemelj result to the matrix case (Appendix~\ref{app:sokhotski_plemelj}).
We then claim that the final Floquet distribution is identical to Eq.~\eqref{eq:floquet_distribution_diagonal}. 

Although the NEGF formalism can also be applied to noninteracting bosonic systems, the present analytical Floquet distribution cannot be directly generalized by replacing the Fermi-Dirac function with the Bose-Einstein function under the same featureless-bath assumption. 
A bosonic particle reservoir in grand-canonical equilibrium requires its chemical potential to lie below the single-particle spectrum, which is incompatible with the unbounded spectrum of the ideal featureless bath. 
Physically consistent bosonic extensions therefore require a reservoir with an appropriate spectral bound, and the resulting Floquet occupations generally depend on the bath spectral function and stability conditions. 
Bosonic Floquet statistics with harmonic-oscillator heat baths have been investigated using Floquet-Markov approaches, for example in Ref.~\cite{liu_classification_2015}. 

\section{DC Floquet Landauer formula}

Having established the steady-state Floquet distribution, we now show how it modifies one of the most fundamental transport formulas.
We extend the Landauer framework to periodically driven systems by formulating a Floquet version of the current expression. 
The setup now consists of a central system coupled to two baths, denoted left (L) and right (R), following standard convention, although the actual spatial direction is irrelevant. 
This is similar to the setup used in Ref.~\cite{camalet_current_2003}, except that their baths are only connected to the two boundaries of the system. 
The total Hamiltonian is block‑structured as follows,
\begin{equation}
    H(t) 
    = \begin{pmatrix}
        H_L & H_{LS} & 0 \\
        H_{SL} & H_S(t) & H_{SR} \\
        0 & H_{RS} & H_R
    \end{pmatrix} .
\end{equation}
The total Green's function can be partitioned similarly. 
The particle current flowing from the left reservoir into the system can be obtained by considering the continuity equation via the Heisenberg equation of motion.
Introducing the projector operator onto the left bath subspace $I_L$ (which corresponds to the number operator in second quantization), the particle current operator reads as 
\begin{equation}
    J_{S \leftarrow L} 
    = -\frac{1}{i\hbar} [I_L, H] 
    = \frac{1}{i\hbar}
    \begin{pmatrix}
        0 & -H_{LS} & 0 \\
        H_{SL} & 0 & 0 \\
        0 & 0 & 0
    \end{pmatrix} .
\end{equation}
By symmetry, the expression for the current exiting the right bath can be simply obtained by changing the labels $L \to R$.
It therefore suffices to present the derivation for one side. 

Within the NEGF formalism, the expectation value of the current operator is expressed in terms of the lesser Green’s function,
\begin{align}
    \langle J_{S \leftarrow L} \rangle(t) 
    &= - i\hbar \Tr_t[ \bm{G}^< \bm{J}_{S \leftarrow L} ] \notag \\
    &= 2 \Re \Tr_t[ \bm{G}_{SL}^< \bm{H}_{LS} ] \notag \\
    &= 2 \Re \Tr_t[ \bm{G}_S^R \bm{\Sigma}_L^< + \bm{G}_S^< \bm{\Sigma}_L^A] , \label{eq:current_expectation}
\end{align}
where the new notation for the trace evaluated at time $t$ is defined as 
\begin{equation}
    \Tr_t \bm{A}(\omega) = \Tr A(t,t) . 
\end{equation}
Note that the trace is incomplete as the variable $t$ is not integrated out, so $\Tr_t \bm{A}(\omega) \bm{B}(\omega) \neq \Tr_t \bm{B}(\omega) \bm{A}(\omega)$ in general, but $\Tr_t \bm{A}(\omega) \bm{P} = \Tr_t \bm{P} \bm{A}(\omega)$ still holds.
This justifies why the current operator can be placed either on the left or right of $\bm{G}^<$ in Eq.~\eqref{eq:current_expectation}, as it is local in time. 
Another useful property is
\begin{equation}
    \Tr_t (\bm{A}(\omega) + \bm{A}(\omega)^\dagger) = 2 \Re \Tr_t \bm{A}(\omega) . \label{eq:tracet_real_hermitian}
\end{equation}

For a two-terminal problem, the system's Green’s functions take the form analogous to Eqs.~\eqref{eq:gr_general} and \eqref{eq:keldysh}, 
\begin{subequations}
\begin{align}
    \bm{G}_S^R &= (\hbar\bm{\Omega} - \bm{H}_S - (\bm{\Sigma}_L^R + \bm{\Sigma}_R^R))^{-1} , \\
    \bm{G}_S^< &= \bm{G}_S^R (\bm{\Sigma}_L^< + \bm{\Sigma}_R^<) \bm{G}_S^A . \label{eq:keldysh_two}
\end{align}
\end{subequations}
As before, the baths are taken to be featureless, where the self-energies are 
\begin{subequations}
\begin{align}
    \Sigma_{L/R}^R &= -i \Gamma_{L/R}/2 , \\
    \Sigma_{L/R}^A &= i \Gamma_{L/R}/2 , \\
    \Sigma_{L/R}^< &= i F_{L/R} \Gamma_{L/R} .  
\end{align}
\end{subequations}
The Fermi functions are required to differ ($F_L \neq F_R$) to establish a gradient for current flow. 
We also take this opportunity to introduce the notation $f_{L/R}$ for the scalar version of the Fermi function for the left/right bath, respectively. 
In general, we also do not need to assume identical baths and therefore denote their spectral functions separately as $\Gamma_L$ and $\Gamma_R$. 

The Meir-Wingreen form of the current expression, Eq.~\eqref{eq:current_expectation}, matches Refs.~\cite{stefanucci_time-dependent_2008} and \cite{mosallanejad_floquet_2024} exactly, and agrees with Ref.~\cite{wu_floquetgreens_2008} up to a factor of two attributable to spin degeneracy.
In particular, Ref.~\cite{mosallanejad_floquet_2024} also employs the Floquet-NEGF formalism to evaluate particle occupations and transport currents, whereas our treatment focuses on obtaining explicit analytical expressions.
The content of the trace can be rearranged as
\begin{align}
    &\qquad \bm{G}_S^R \bm{\Sigma}_L^< + \bm{G}_S^< \bm{\Sigma}_L^A \notag \\
    &= \bm{G}_S^R \bm{\Sigma}_L^< \bm{G}_S^A (\hbar\bm{\Omega} - \bm{H}_S -\bm{\Sigma}_R^A) + \bm{G}_S^R \bm{\Sigma}_R^< \bm{G}_S^A \bm{\Sigma}_L^A \notag \\
    &= \bm{G}_S^R \bm{\Sigma}_L^< \bm{G}_S^A (\hbar\bm{\Omega} - \bm{H}_S) \notag \\
    &\quad - \bm{G}_S^R \bm{\Sigma}_L^< \bm{G}_S^A \bm{\Sigma}_R^A
    + \bm{G}_S^R \bm{\Sigma}_R^< \bm{G}_S^A \bm{\Sigma}_L^A . \label{eq:kadanoff_baym_left}
\end{align}
Incidentally, the one-bath version of Eq.~\eqref{eq:kadanoff_baym_left} corresponds to the Kadanoff-Baym equation of motion for $\bm{G}_S^<$. 
We shall see that the first term on the right‑hand side of Eq.~\eqref{eq:kadanoff_baym_left} yields the AC part of the current, whereas the second and third terms yield the DC part, with all contributions being first order in $\Gamma_L+\Gamma_R$.

Although an analytical expression for the AC current is difficult to obtain, the DC current admits a surprisingly simple formula in the weak coupling limit.
The DC current can be expressed in terms of the couplings as
\begin{equation}
    \langle J_{S \leftarrow L} \rangle_{\mathrm{DC}} 
    = \Re \Tr_t [\bm{G}_S^R \bm{F}_L \bm{\Gamma}_L \bm{G}_S^A \bm{\Gamma}_R 
    - \bm{G}_S^R \bm{F}_R \bm{\Gamma}_R \bm{G}_S^A \bm{\Gamma}_L ] . \label{eq:current_caroli}
\end{equation}
In the absence of driving, the Floquet matrices are block diagonal and commute, recovering the standard Caroli formula for steady‑state transport~\cite{caroli_direct_1971,haug_quantum_2008,datta_mesoscopic_1995,wang_transport_2023}, where the current is given by the transmission probability multiplied by the occupation difference, $(f_L - f_R)$. 

In the presence of driving, the Floquet Fermi functions in Eq.~\eqref{eq:current_caroli} cannot be easily taken outside. 
Observe that $\bm{G}^<_S$ (to zeroth order in coupling) can be expressed as 
\begin{equation}
    \bm{G}_S^< = 2 \pi i \bm{P} \bar{\bm{\rho}} \delta(\hbar\bm{\Omega} - \bm{H}_F) \bm{P}^\dagger , \label{eq:gless_twobath_zeroth}
\end{equation}
which is a matrix version of Eq.~\eqref{eq:gless_featureless_components2} generalized to two baths. 
The steady-state distribution for the system in the two-bath case is a weighted average,
\begin{equation}
    \bar{\rho} = \frac{\Gamma_L \rho_L + \Gamma_R \rho_R}{\Gamma_L + \Gamma_R} ,
\end{equation}
with $ \rho_{L/R}$ being the Floquet distribution defined as if the system is only connected to the corresponding bath. 
The factorization used in Eq.~\eqref{eq:gless_twobath_zeroth} together with the assumption that the spectral function universally commutes gives
\begin{align}
    &\quad \bm{G}_S^R \bm{F}_L \bm{\Gamma}_L \bm{G}_S^A \bm{\Gamma}_R - \bm{G}_S^R \bm{F}_R \bm{\Gamma}_R \bm{G}_S^A \bm{\Gamma}_L \notag \\
    &= 2 \pi \frac{\Gamma_L \Gamma_R}{\Gamma_L + \Gamma_R} 
    \bm{P} (\bm{\rho}_L - \bm{\rho}_R) \delta(\hbar\bm{\Omega} - \bm{H}_F) \bm{P}^\dagger .
\end{align}
What remains is to transform back to the time domain, which is similar to going from $-i \hbar \bm{G}_S^< \to \rho_S(t)$.
The outermost $P(t)$ and $P(t)^\dagger$ cancel by the cyclic property of the internal trace, resulting in a time-independent quantity, 
\begin{equation}
    \langle J_{S \leftarrow L} \rangle_{\mathrm{DC}}
    = \frac{1}{\hbar} \frac{\Gamma_L \Gamma_R}{\Gamma_L + \Gamma_R} \Tr 
    (\bm{\rho}_L - \bm{\rho}_R) . \label{eq:floquet_landauer}
\end{equation}
Remarkably, the final result differs from the non-driven case simply by replacing the Fermi distribution ($f_{L/R}$) with the Floquet distribution ($\rho_{L/R}$), and only the diagonal elements corresponding to occupation contribute. 
The harmonic mean of the couplings reflects the effective conductance of two conductors arranged in series.
As a consistency check, if the two baths have the same initial temperature but different chemical potentials, say $\mu_L > \mu_R$, then current flows from the left bath to the right through the system without accumulation. 
Increasing driving strength generally, though not invariably, leads to a reduction in current flow. 

\begin{figure}[htb]
    \centering
    \includegraphics[width=\linewidth]{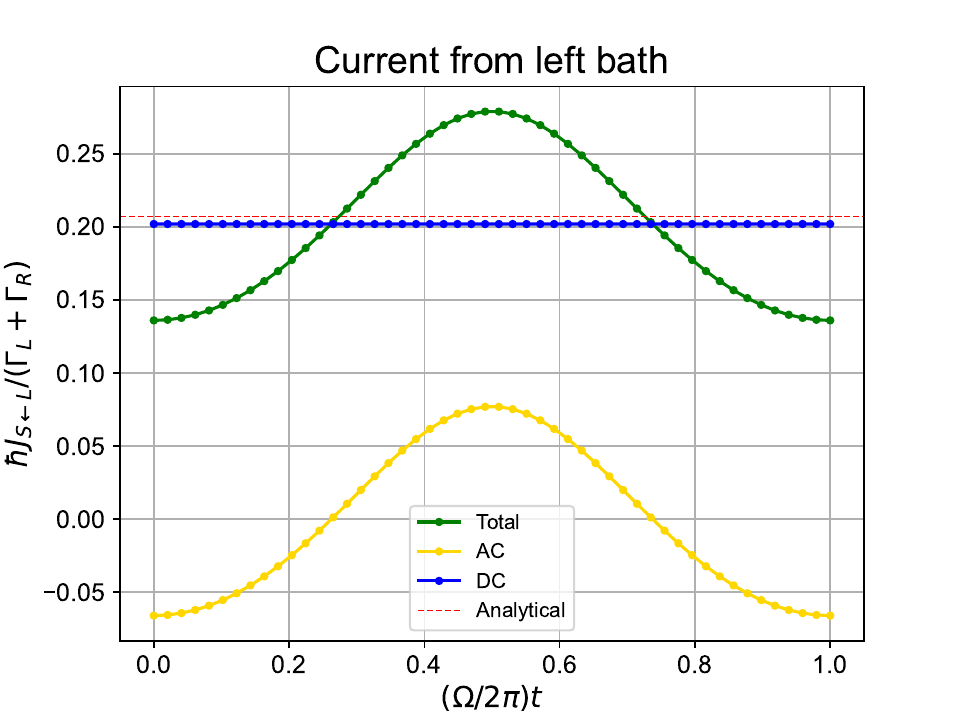}
    \caption{
    \label{fig:current_numerical} 
    Numerical computation of the current exiting from the left bath over one period using Eq.~\eqref{eq:current_expectation}. 
    To show leading-order behaviour, the current is normalized by $\Gamma_L + \Gamma_R$. 
    The same sinusoidal resonant level model is used, with the Hamiltonian given by Eq.~\eqref{eq:sinusoidal_hamiltonian}. 
    The Hamiltonian parameters are $\hbar\Omega = \SI{2.07}{\electronvolt}$, $\epsilon_0 = \SI{0}{\electronvolt}$, and $\epsilon_1 = \SI{372}{\milli\electronvolt}$. 
    Both baths are kept at the same temperature of $\SI{300}{\kelvin}$, but the chemical potentials differ, with $\mu_L = \SI{82.7}{\milli\electronvolt}$ and $\mu_R = -\SI{82.7}{\milli\electronvolt}$. 
    The bath couplings are set slightly differently, with $\Gamma_L = 0.001 \hbar\Omega$ and $\Gamma_R = 0.0015 \hbar\Omega$, although the results remain robust under variations. 
    The ``Total'', ``AC'', and ``DC'' currents are computed numerically from their respective terms in Eq.~\eqref{eq:kadanoff_baym_left}, while the analytical DC level is obtained from Eq.~\eqref{eq:floquet_landauer}. 
    In the implementation, all Floquet matrices are truncated to size $121 \times 121$. 
    This example demonstrates that the AC and DC components of the current can be of comparable magnitude, and verifies that the average of the AC component is zero. 
    }
\end{figure}

Now we turn to the AC part of the current, 
\begin{equation}
    \langle J_{S \leftarrow L} \rangle_{\mathrm{AC}}(t) 
    = 2 \Re \Tr_t[ \bm{G}_S^R \bm{\Sigma}_L^< \bm{G}_S^A (\hbar\bm{\Omega} - \bm{H}_S) ] . \label{eq:ac_current}
\end{equation}
Observe that $\bm{G}_S^R \bm{\Sigma}_L^< \bm{G}_S^A$ (to zeroth order) admits a factorization similar to $\bm{G}_S^<$ in Eq.~\eqref{eq:gless_twobath_zeroth}, and thus 
\begin{align}
    &\quad \Re \Tr_t[\bm{G}_S^R \bm{\Sigma}_L^< \bm{G}_S^A (\hbar\bm{\Omega} - \bm{H}_S)]\notag \\
    &= \Re \Tr_t \biggl[2 \pi i \frac{\Gamma_L}{\Gamma_L + \Gamma_R} \bm{\rho}_L \delta(\hbar\bm{\Omega} - \bm{H}_F) (\hbar\bm{\Omega} - \bm{H}_F) \biggr] , \label{eq:ac_delta}
\end{align}
which evaluates to zero, since $x \delta(x) = 0$ in the sense of distributions. 
It follows from Eq.~\eqref{eq:tracet_real_hermitian} that there is no AC contribution at the zeroth order. 
The leading contribution is first order in the coupling, and the numerical results in Fig.~\ref{fig:current_numerical} show that the AC current can be comparable in magnitude to the DC current. 
The numerical current is evaluated directly using the conventional Floquet-matrix NEGF approach, independently of the analytical Floquet distribution used to obtain Eq.~\eqref{eq:floquet_landauer}. 
The close agreement between the numerical and analytical results therefore serves as an additional consistency check of our Floquet-Landauer formula.

Although an analytical formula for the AC current is complicated (Appendix~\ref{app:analytical_ac}), its time average can be shown to be exactly zero by considering the terms $\hbar\bm{\Omega}$ and $\bm{H}_S$ separately. 
Since $\bm{H}_S$ is local, an anti-Hermiticity argument shows that $\Re \Tr_t(\bm{G}_S^R \bm{\Sigma}_L^< \bm{G}_S^A \bm{H}_S) = 0$ exactly. 
For the $\hbar\bm{\Omega}$ term, since we are only concerned with the DC component, it suffices to consider only the block-diagonal part of $\bm{G}_S^R \bm{\Sigma}_L^< \bm{G}_S^A$, which commutes with $\bm{\Omega}$. 
(The diagonal blocks of $\bm{\Omega}$ are proportional to the identity.) 
By the same anti-Hermiticity argument, we get $\Re \Tr_t(\bm{G}_S^R \bm{\Sigma}_L^< \bm{G}_S^A \bm{\Omega}) = 0$.
Therefore, Eq.~\eqref{eq:floquet_landauer} represents the sole contribution to the DC current in the weak-coupling limit, as corroborated by numerical evidence in Fig.~\ref{fig:current_numerical}.

\section{Conclusion}

In this work, we have derived an analytical Floquet steady-state distribution for periodically driven quantum systems coupled to baths using the NEGF formalism. 
By recognizing the role of the micromotion operator in the diagonalization process, factorized expressions for the retarded, advanced, and lesser Green’s functions were derived assuming weak system–bath coupling. 
This led to an analytical expression for the reduced density matrix, revealing the general Floquet Fermi distribution that governs steady-state population.
Periodic driving reshapes the equilibrium Fermi–Dirac statistics into a Floquet-shifted weighted sum, with well-defined weights determined by the micromotion operator. 
As an application, the Landauer formula is generalized by replacing the equilibrium Fermi function with its Floquet-modified counterpart. 
Taken together, the present framework establishes a direct connection between Floquet engineering and nonequilibrium quantum statistics, providing a unified framework for describing periodically driven open quantum systems.

Beyond the sinusoidal resonant-level example used in numerical calculations, the formalism is general and applicable to arbitrary periodic drives in multi-level systems and to diverse bath spectral functions. 
The Floquet representation and convolution theorem offer computational tractability, remaining applicable even in contexts beyond this study, where analytical solutions are infeasible.
Future directions include extending the framework to interacting systems, further exploring non-Markovian baths, and applying the formalism to experimentally relevant platforms. 
We anticipate that the methods presented here will serve as a useful basis for both theoretical investigations and practical implementations of Floquet‑engineered quantum systems.

\section{Acknowledgements}
G.T. is supported by Science Challenge Project (Grant No. TZ2025017) and National Natural Science Foundation of China (Grant No. 12374048). 

\appendix
\section{Fourier transform of Lorentzian-weighted Fermi function} \label{app:lerch_digamma}

Here, we show how to transform Eq.~\eqref{eq:gless_featureless_components} back to the time domain exactly using a generalization of Eq.~\eqref{eq:back_equal_time} for unequal times, 
\begin{align}
    G(t, t') 
    = \sum_{m} e^{-i m \Omega t} \int_{-\infty}^{\infty} \frac{\d \omega}{2\pi} \, e^{-i \omega (t-t')} G_{m 0}(\omega) .
\end{align}
For convenience, let $z_a = [H_F]_{aa}-m\hbar\Omega-i\frac{\Gamma}{2}$, $z_b = [H_F]_{bb}-n\hbar\Omega+i\frac{\Gamma}{2}$ and $t_r = t-t'$.
We want to evaluate the integral
\begin{align}
    \mathcal{I}(t_r) 
    = \int_{-\infty}^{\infty} \d E \, e^{\frac{E t_r}{i \hbar}} 
    f(E) \frac{\Gamma/2}{(E-z_a)(E-z_b)} ,
\end{align}
where $f$ is the Fermi function and assuming $\Gamma>0$.

Integrals involving a Fermi function and a single complex pole have previously been expressed in terms of the Hurwitz--Lerch transcendent~\cite{ridley_calculation_2016}. 
Here, we extend this approach to the product of retarded and advanced Green's function denominators, using contour integration to evaluate the inverse Fourier transform. 
The integrand has two families of poles: the Lorentzian poles at $z_a$ and $z_b$, and the Matsubara poles of the Fermi function. 
For $t_r>0$, the contour is closed in the lower half-plane; for $t_r<0$, it is closed in the upper half-plane. 
The integral can be evaluated exactly, and the two cases corresponding to the sign of $t_r$ can be combined.
Assuming $t_r\neq 0$, we get 
\begin{align}
    \mathcal{I}(t_r)
    &= \frac{\Gamma/2}{z_b-z_a}
    \bigg(
        2\pi i\,e^{-iz_st_r/\hbar}f(z_s) - e^{-i\mu t_r/\hbar}  \notag \\
        &\quad\times \sqrt{q}
        \left[
            \Phi(q,1,v_b^{(s)}) - \Phi(q,1,v_a^{(s)})
        \right]
    \bigg) . \label{eq:lerch_expression}
\end{align}
Here, we define
\begin{subequations}
\begin{align}
    s &= \operatorname{sgn}(t_r) , \\
    z_s &= \frac{(1+s) z_a + (1-s) z_b}{2} , \\
    q &= e^{-2\pi|t_r|/(\beta\hbar)} , \\
    v_j^{(s)} &= \frac{1}{2} + s \frac{i \beta(\mu-z_j)}{2\pi}, \qquad j=a,b .
\end{align}
\end{subequations}
Due to the argument being complex, $f(z_s)$ is the meromorphic continuation of the Fermi function. 
The function $\Phi(q,1,v)$ is the Lerch transcendent,
\begin{align}
    \Phi(q,1,v) = \sum_{n=0}^{\infty}\frac{q^n}{n+v} .
\end{align}
The first term arises from the poles $z_a$ or $z_b$ of the Lorentzian denominator, and the second term arises from the Matsubara poles of the Fermi function and is expressed through the Lerch transcendent.
When a Lorentzian pole coincides with a Matsubara pole, Eq.~\eqref{eq:lerch_expression} is understood in the limiting sense.

In the limit $t_r\to 0$, we have $q\to 1^-$.
Although the individual Lerch transcendents diverge, their difference remains finite,
\begin{align}
    \lim_{q\to 1^-}
    \left[
        \Phi(q,1,v_b) - \Phi(q,1,v_a)
    \right]
    = \Psi(v_a)-\Psi(v_b) .
\end{align}
This follows from the series representation of the digamma function,
\begin{align}
    \Psi(v)
    = - \gamma + \sum_{n=0}^{\infty}
    \left(\frac{1}{n+1} - \frac{1}{n+v}\right) .
\end{align}
We then get
\begin{align}
    \mathcal{I}(0^+) 
    = \frac{\Gamma/2}{z_b - z_a}
    \left[
        2\pi i f(z_a) 
        + \Psi\left(v_b^{(+)}\right) 
        - \Psi\left(v_a^{(+)}\right)
    \right] .
\end{align}
The Fermi function can be eliminated using the digamma reflection formula,
\begin{align}
    i \pi [1 - 2 f(z_j)]
    &= \pi\cot(\pi v_j^{(+)}) \notag \\
    &= \Psi\left(1-v_j^{(+)}\right)-\Psi\left(v_j^{(+)}\right) .
\end{align}
The Lerch-transcendent expression reduces to
\begin{align}
    \mathcal{I}(0^+)
    &= \frac{\Gamma/2}{z_b-z_a} 
    \left[
        i\pi + \Psi\left(v_b^{(+)}\right) - \Psi\left(1 - v_a^{(+)}\right)
    \right] \notag \\
    &= \frac{i\pi \Gamma/2}{\epsilon_b-\epsilon_a+i\Gamma}
    \left[
        f_{\Gamma/2}(\epsilon_a) + f_{\Gamma/2}(\epsilon_b)^*
    \right] , \label{eq:digamma_expression}
\end{align}
with $\epsilon_a = [H_F]_{aa} - m\hbar\Omega$ and $\epsilon_b = [H_F]_{bb} - n\hbar\Omega$ being the Floquet-shifted quasienergies. 
The complex auxiliary Fermi function $f_{\Gamma/2}$ was defined in Eq.~\eqref{eq:fermi_auxiliary}, and we use this digamma expression to obtain Eq.~\eqref{eq:distribution_exact}. 

The limit $t_r \to 0^-$ can be evaluated analogously to arrive at the same results. 
Alternatively, the equality of the two one-sided limits follows directly from the dominated convergence theorem. 
Indeed, the integrand decays as $E^{-2}$ for $E\to-\infty$ and exponentially for $E\to+\infty$, ensuring absolute integrability. 
Moreover, since $|e^{-iEt_r/\hbar}|=1$, the absolute value of the integrand is independent of $t_r$ and thus provides an integrable dominating function. 
The dominated convergence theorem therefore allows the limit $t_r \to 0$ to be interchanged with the energy integral, establishing the continuity of $\mathcal{I}(t_r)$ at $t_r=0$. Consequently, $\mathcal{I}(0^+) = \mathcal{I}(0^-) = \mathcal{I}(0)$.
\section{Keldysh equation in time domain} \label{app:keldysh_time}

Here, the Floquet distribution is alternatively derived starting from the Keldysh equation in the time domain,  
\begin{align}
    G_S^<(t, t') &= \int_{-\infty}^{\infty} \d t_1 \int_{-\infty}^{\infty} \d t_2 \, \notag \\
    &\quad\times G_S^R(t, t_1) \Sigma^<(t_1, t_2) G_S^A(t_2, t') . \label{eq:keldysh_eqn_time}
\end{align}
A merit of this approach is that it retains the full time dependence of the system, despite the final distribution being time independent.
The expressions for the Green’s functions in the time domain may also prove useful in other applications. 

For the infinite featureless bath discussed earlier, with self-energy given by Eq.~\eqref{eq:general_sigmar}, the ingredients for the Keldysh equation are
\begin{subequations}
\begin{align}
    G_S^R(t,t_1) &= \frac{1}{i\hbar} \Theta(t-t_1) U_S(t) U_S(t_1)^\dagger e^{- \frac{\Gamma}{2\hbar} (t-t_1)} , \\
    \Sigma^<(t_1, t_2) &= i \Gamma f(t_1 - t_2) , \\    
    G_S^A(t_2, t') &= \frac{-1}{i\hbar} \Theta(t'-t_2) U_S(t_2) U_S(t')^\dagger e^{\frac{\Gamma}{2\hbar} (t_2-t')} .
\end{align}
\end{subequations}
Substituting into Eq.~\eqref{eq:keldysh_eqn_time}, we have 
\begin{align}
    &\quad U_S(t)^\dagger G_S^<(t, t') U_S(t') e^{\frac{\Gamma}{2\hbar} (t+t')} \notag \\
    &= \frac{i\Gamma}{\hbar^2} \int_{-\infty}^{t} \d t_1 \int_{-\infty}^{t'} \d t_2 \, e^{\frac{\Gamma}{2\hbar} (t_1 + t_2)} \notag \\
    &\quad\times U_S(t_1)^\dagger f(t_1 - t_2) U_S(t_2) .
\end{align}
The following change of integration variables is applied, $t_r = t_1 - t_2$, $t_a = \frac{1}{2} (t_1 + t_2)$. 
Consequently, the boundaries of integration have to be adjusted, 
\begin{align}
    &\quad U_S(t)^\dagger G_S^<(t, t') U_S(t') e^{\frac{\Gamma}{2\hbar} (t+t')} \notag \\
    &= \frac{i\Gamma}{\hbar^2} \int_{-\infty}^{\infty} \d t_r \int_{-\infty}^{\frac{1}{2}(t + t' - |t_r - t + t'|)} \d t_a \, e^{\frac{\Gamma}{\hbar} t_a} \notag \\
    &\quad\times U_S(t_a + t_r/2)^\dagger f(t_r) U_S(t_a - t_r/2) . \label{eq:change_coordinates}
\end{align}
To proceed, the form of $U_S$ is taken from Floquet's theorem, and the Fourier series representation of the periodic micromotion operator $P(t)$ is used. 
Eq.~\eqref{eq:change_coordinates} is thus 
\begin{align}
    &\quad U_S(t)^\dagger G_S^<(t, t') U_S(t') e^{\frac{\Gamma}{2\hbar} (t+t')} \notag \\
    &= \frac{i\Gamma}{\hbar^2} \int_{-\infty}^{\infty} \d t_r \int_{-\infty}^{\frac{1}{2}(t + t' - |t_r - t + t'|)} \d t_a \, e^{\frac{\Gamma}{\hbar} t_a} \notag \\
    &\quad\times f(t_r) \sum_{nm} e^{i n \Omega (t_a + t_r/2)} e^{-i m \Omega (t_a - t_r/2)} \notag \\
    &\quad\times e^{-\frac{t_a + t_r/2}{i\hbar} H_F} P_n^\dagger P_m e^{\frac{t_a - t_r/2}{i\hbar} H_F} \notag \\
    &= \frac{i}{\hbar} \int_{-\infty}^{\infty} \d t_r \mathcal{I}(t_r) f(t_r) ,
\end{align} 
where the inner integral involving $t_a$ has the following matrix elements: 
\begin{align}
    \mathcal{I}(t_r)_{ab} 
    = \frac{\Gamma}{\hbar} \int_{-\infty}^{\frac{1}{2}(t + t' - |t_r - t + t'|)} \d t_a \, e^{\frac{\Gamma}{\hbar} t_a} \sum_{nm} [P_n^\dagger P_m]_{ab} \notag \\
    \times e^{i \left((n-m) \Omega + \frac{[H_F]_{aa} - [H_F]_{bb}}{\hbar} \right) t_a} e^{i \left(\frac{m+n}{2} \Omega + \frac{[H_F]_{aa} + [H_F]_{bb}}{2\hbar} \right) t_r} .
\end{align}

Next, the weak bath coupling limit $\Gamma \to 0^+$ is taken.
We use the following fact,
\begin{align}
    \frac{\Gamma}{\hbar} \int_{-\infty}^T \d t_a \, e^{(\Gamma/\hbar + i\omega) t_a} 
    &= \frac{\Gamma}{\Gamma + i \hbar \omega} e^{(\Gamma/\hbar + i\omega) T} \notag \\
    &\xrightarrow{\Gamma \to 0^+}
    \begin{cases}
        1, \quad &\text{if } \omega = 0 , \\
        0, \quad &\text{if } \omega \neq 0 ,
    \end{cases} \label{eq:infinitesimal_damping_identity}
\end{align}
for $T < \infty$ and fixed $\omega$. 
Thus, in the nondegenerate case, we immediately get $\mathcal{I}(t_r)_{ab} = 0$. 
In the degenerate case, where $[H_F]_{aa} - [H_F]_{bb} = k \hbar \Omega$, only terms with $m-n = k$ in the summation survive the integral,
\begin{equation}
    \mathcal{I}(t_r)_{ab} 
    = \sum_{n} e^{i \left(n \Omega + \frac{[H_F]_{aa}}{\hbar} \right) t_r}  [P_{n}^\dagger P_{n+k}]_{ab} .
\end{equation}

The remaining integral only involves $t_r$. 
The damping factor $e^{\frac{\Gamma}{2\hbar}(t+t')}$ can be neglected to zeroth order, and there is no remaining dependence on $\Gamma$, 
\begin{align}
    &\quad [U_S(t)^\dagger G_S^<(t, t') U_S(t')]_{ab} \notag \\
    &= \frac{i}{\hbar} \int_{-\infty}^{\infty} \d t_r \, f(t_r) 
    \sum_{n} e^{i \left(n \Omega + \frac{[H_F]_{aa}}{\hbar} \right) t_r}  [P_{n}^\dagger P_{n+k}]_{ab} .
\end{align} 
The evaluation of the remaining $t_r$ integral is a usual Fourier transform, \begin{align}
    &\quad [U_S(t)^\dagger G_S^<(t, t') U_S(t')]_{ab} \notag \\
    &= \frac{-1}{i \hbar} \sum_{n} f([H_F]_{aa} + n \hbar \Omega) [P_{n}^\dagger P_{n+k}]_{ab} .
\end{align} 

Hence, the precursor to the zeroth-order Floquet distribution in the main text, Eq.~\eqref{eq:gless_equal_time}, is recovered in the time domain. 
\section{Sokhotski-Plemelj generalization} \label{app:sokhotski_plemelj}

This appendix justifies why replacing the featureless bath with a frequency-dependent spectral function leaves the weak-coupling Floquet distribution unchanged. 
The original Sokhotski-Plemelj formula states
\begin{equation}
    \frac{\Gamma}{\omega^2 + \Gamma^2} \to \pi \delta(\omega) ,
\end{equation}
in the limit $\Gamma \to 0^+$. 
This can be generalized by allowing $\Gamma$ to vary, provided it remains small. 
Specifically, we set $\Gamma(\omega) = \eta k(\omega)$, where $\eta$ is a scaling parameter, and $k(\omega)$ is an envelope function. 
We assume $k(\omega)$ to be continuous and bounded, which ensures that $\Gamma(\omega)$ vanishes as $\eta \to 0^+$. 
We also require $k(0) > 0$ to reproduce the singularity at $x=0$. 
Under these conditions, we claim
\begin{equation}
    \frac{\Gamma(\omega)}{\omega^2 + \Gamma(\omega)^2} \to \pi \delta(\omega) .
\end{equation}

To establish the result, we need to show that for any test function $h(\omega)$,
\begin{equation}
    \int_a^b \d \omega \, \frac{\Gamma(\omega)}{\omega^2 + \Gamma(\omega)^2} h(\omega) 
    =
    \begin{cases}
        \pi h(0) , & a < 0 < b , \\
        0 , & \text{otherwise} .
    \end{cases}
\end{equation}
In a small neighborhood around $\omega = 0$, the continuity of $k(\omega)$ implies $\Gamma(\omega) \approx \eta k(0)$. 
Applying the Sokhotsky-Plemelj formula,
\begin{equation}
    \frac{\eta k(0)}{\omega^2 + \eta^2 k(0)^2} \to \pi \delta(\omega) .
\end{equation}
Away from $\omega=0$, we note that  
\begin{equation}
    \frac{\Gamma(\omega)}{\omega^2 + \Gamma(\omega)^2} < \frac{\Gamma(\omega)}{\omega^2} \to 0 ,
\end{equation}
which completes the proof. 
Thus, $\Gamma$ is allowed to depend on $\omega$, provided that it remains small and satisfies $k(0) > 0$.

Next, we rewrite Eq.~\eqref{eq:sokhotsky_plemlj_variant} in terms of the self-energies.
Let $\Sigma^R = \Lambda - i \Gamma/2$, $\Sigma^A = \Lambda + i \Gamma/2$. 
Then, the Sokhotsky-Plemelj formula for two variables is 
\begin{align}
    \frac{\Sigma^R - \Sigma^A}{(x - \Sigma^R)(y - \Sigma^A)} 
    = 
    \frac{-i \Gamma}{(x - \Lambda + i \Gamma/2)(y - \Lambda - i \Gamma/2)} \notag \\
    \overset{\Gamma \to 0^+}{\rightarrow} -2i \times 
    \begin{cases}
        \pi \delta(x-\Lambda) , \quad &\text{if } x = y , \\
        \pi \delta(y-\Lambda) - i \mathcal{P}(\frac{1}{y-\Lambda}) , \quad &\text{if } x-\Lambda = 0 , \\
        \pi \delta(x-\Lambda) + i \mathcal{P}(\frac{1}{x-\Lambda}) , \quad &\text{if } y-\Lambda = 0 , \\
        0 , \quad &\text{otherwise} ,
    \end{cases} 
\end{align} 
with $\mathcal{P}$ denoting the Cauchy principal value. 
Ultimately, $\Sigma^R$ must be small, so the limit $\Lambda \to 0$ is taken. 
Note that the limits $\Lambda \to 0$ and $\Gamma \to 0^+$ can be taken in any order. 

We now generalize to the matrix case, with $\Sigma^A = (\Sigma^R)^\dagger$.
In this setting, both $\Gamma = i (\Sigma^R - \Sigma^A)$ and $2 \Lambda = \Sigma^R + \Sigma^A$ are Hermitian matrices. 
Our goal is to evaluate 
\begin{equation}
    (X - \Sigma^R)^{-1} (\Sigma^R - \Sigma^A) (Y - \Sigma^A)^{-1} ,
\end{equation}
where $X$ and $Y$ are now matrix quantities. 
The matrix inverse can be evaluated according to Cramer's rule,
\begin{equation}
    (X - \Sigma^R)^{-1} = \frac{1}{\det(X - \Sigma^R)} \adj(X - \Sigma^R) .
\end{equation}
Since $\Gamma$ is small, there is no problem approximating $\adj(X - \Sigma^R) \approx \adj(X - \Lambda)$, as the entries of the adjugate matrix are continuous polynomials. 
Meanwhile, the determinant expands as
\begin{align}
    &\det(X - \Sigma^R) = \det(X - \Lambda) \notag \\
    &\times [1 + \Tr(i (X-\Lambda)^{-1} \Gamma/2) + \mathcal{O}(\|\Gamma\|^2)] .
\end{align}
Combining these results gives
\begin{equation}
    (X - \Sigma^R)^{-1} \approx \frac{(X-\Lambda)^{-1}}{1 + \frac{i}{2} \Tr(\Gamma (X-\Lambda)^{-1})} ,
\end{equation}
which is the desired form for applying the Sokhotsky-Plemelj formula. 
 
Without loss of generality, we work in the basis where $\Sigma^R - \Sigma^A$ is diagonal and purely imaginary, 
\begin{equation}
    \Sigma^R - \Sigma^A = - i 
    \begin{pmatrix}
        \Gamma_1 & & \\
        & \ddots & \\
        & & \Gamma_N 
    \end{pmatrix} .
\end{equation}
The matrix product can then be evaluated element-wise,
\begin{align}
    &\quad [(X-\Sigma^R)^{-1} (\Sigma^R - \Sigma^A) (Y - \Sigma^A)^{-1}]_{ac} \notag \\
    &= \sum_b \frac{(X-\Lambda)^{-1}_{ab}}{1 + i \sum_i (X-\Lambda)^{-1}_{ii} \Gamma_i/2}  \notag \\
    &\quad\times (-i \Gamma_b) \frac{(Y-\Lambda)^{-1}_{bc}}{1 + i \sum_j (Y-\Lambda)^{-1}_{jj} \Gamma_j/2} .
\end{align}
In the limit $\Gamma \to 0$, only the terms with $i=b$ and $j=b$ are important and need to be kept in the denominators, giving
\begin{align}
    &\quad [(X-\Sigma^R)^{-1} (\Sigma^R - \Sigma^A) (Y - \Sigma^A)^{-1}]_{ac} \notag \\
    &= - i \sum_b \frac{(X-\Lambda)^{-1}_{ab} \Gamma_b (Y-\Lambda)^{-1}_{bc}}{[1 + i (X-\Lambda)^{-1}_{bb} \Gamma_b/2][1 + i (Y-\Lambda)^{-1}_{bb} \Gamma_b/2]} .
\end{align}
Now, directly applying the Sokhotski-Plemelj formula yields the following,
\begin{align}
    &\quad [(X-\Sigma^R)^{-1} (\Sigma^R - \Sigma^A) (Y - \Sigma^A)^{-1}]_{ac} \notag \\
    &= -2 \pi i \sum_b (X-\Lambda)^{-1}_{ab} \delta((X - \Lambda)_{bb}) (Y-\Lambda)^{-1}_{bc} ,
\end{align}
provided that $X_{bb}$ and $Y_{bb}$ are simultaneously small. 
We assume all $\Gamma_b > 0$; otherwise, sign functions have to be included. 
To qualitatively interpret the results, divergence occurs when the diagonal elements of $X$ and $Y$ are small in the eigenbasis of $\Gamma$. 

As a final step to replicate the Keldysh equation, the Fermi function is included as a middle term.
Assuming that $F$ is diagonal,  
\begin{align}
    &\quad [-(X-\Sigma^R)^{-1} (\Sigma^R - \Sigma^A) F (Y - \Sigma^A)^{-1}]_{ac} \notag \\
    &= 2 \pi i \sum_b (X-\Lambda)^{-1}_{ab} F_{bb} \delta((X - \Lambda)_{bb}) (Y-\Lambda)^{-1}_{bc} .
\end{align}
The result coincides with the standard expression in the case where $\Gamma$ is proportional to the identity. 
Thus, the same Floquet distribution is obtained, regardless of the specifics of $\Gamma(\omega)$. 
We have assumed that $\Gamma$ is positive definite, which is evident considering its definition in Eq.~\eqref{eq:bath_spectral}. 
\section{Analytical AC current} \label{app:analytical_ac}

In this appendix, we derive an analytical expression for the AC contribution to the Floquet-Landauer current and obtain its weak-coupling limit. 
We consider the two-bath setup with featureless reservoirs and total coupling $\Gamma=\Gamma_L+\Gamma_R$.
We begin directly from Eq.~\eqref{eq:ac_current}. 
Using Eq.~\eqref{eq:shirley_diagonalization} and the cyclic property of the equal-time trace for time-local operators, we obtain
\begin{align}
    \langle J_{S\leftarrow L}\rangle_{\mathrm{AC}}(t)
    &= 2\Re\operatorname{Tr}_t
    \Big[ \notag \\
    &\quad (\hbar\bm{\Omega} - \bm{H}_F + i\Gamma/2)^{-1}
    (i\Gamma_L \bm{P}^\dagger\bm{F}_L\bm{P}) \notag \\
    &\quad\times (\hbar\bm{\Omega} - \bm{H}_F - i\Gamma/2)^{-1}
    (\hbar\bm{\Omega} - \bm{H}_F)
    \Big] .
\end{align}

The internal trace sets $b=a$, while the Floquet indices $m$ and $n$ remain independent.
The inverse Floquet transform gives
\begin{widetext}
\begin{align}
    \langle J_{S\leftarrow L} \rangle_{\mathrm{AC}}(t)
    = 2\Re \sum_{amn} e^{-i(m-n)\Omega t}
    &\int_{-\Omega/2}^{\Omega/2} \frac{\d\omega}{2\pi} \,
    \frac{
        i \Gamma_L (\hbar\omega+n\hbar\Omega-[H_F]_{aa})
    }{
        (\hbar\omega + m\hbar\Omega - [H_F]_{aa} + i\Gamma/2)
        (\hbar\omega + n\hbar\Omega - [H_F]_{aa} - i\Gamma/2)
    } \notag \\
    &\times \sum_l f_L(\hbar\omega+l\hbar\Omega)
    [P_{l-m}^\dagger P_{l-n}]_{aa} .
\end{align}
\end{widetext}

The matrix $\bm{G}_S^R \bm{\Sigma}_L^< \bm{G}_S^A$ is anti-Hermitian. 
Taking the complex conjugate and exchanging $m \leftrightarrow n$, the operation $2\Re$ replaces the numerator $(\hbar\omega + n\hbar\Omega-[H_F]_{aa})$ by $(n-m)\hbar\Omega$, eliminating the continuous frequency variable from the numerator.
Introducing $k=m-n$ and $\xi=l-m$, we unfold the Floquet frequency integral onto the entire real axis by setting $E=\hbar\omega+l\hbar\Omega$ and removing the $l$ summation.
The AC current becomes
\begin{widetext}
\begin{align}
    \langle J_{S\leftarrow L}\rangle_{\mathrm{AC}}(t)
    = \sum_{ak\xi} e^{-ik\Omega t}
    \int_{-\infty}^{\infty}
    \frac{\d E}{2\pi\hbar}
    \frac{
        i\Gamma_L (-k\hbar\Omega) f_L(E) [P_\xi^\dagger P_{\xi+k}]_{aa}
    }{
        (E - \xi\hbar\Omega -[H_F]_{aa} + i\Gamma/2)
        (E - (\xi+k)\hbar\Omega -[H_F]_{aa} - i\Gamma/2)
    } . 
\end{align}
\end{widetext}

The remaining integral is the equal-time integral evaluated in Appendix~\ref{app:lerch_digamma}. 
Using Eq.~\eqref{eq:digamma_expression}, we obtain
\begin{widetext}
\begin{align}
    \langle J_{S\leftarrow L}\rangle_{\mathrm{AC}}(t)
    = \frac{\Gamma_L}{2\hbar}
    \sum_{a k \xi} e^{-ik\Omega t}
    \frac{k\hbar\Omega}{k\hbar\Omega + i\Gamma}
    [P_\xi^\dagger P_{\xi+k}]_{aa} 
    \bigl[
        f_{L,\Gamma/2}([H_F]_{aa} + \xi\hbar\Omega) 
        + f_{L,\Gamma/2}([H_F]_{aa} + (\xi+k)\hbar\Omega)^*
    \bigr] . \label{eq:ac_exact_digamma}
\end{align}
\end{widetext}
Here, $f_{L,\Gamma/2}$ denotes the auxiliary Fermi function defined in Eq.~\eqref{eq:fermi_auxiliary}, evaluated using $\beta_L$ and $\mu_L$. 
The broadening entering this function is the total $\Gamma$, rather than $\Gamma_L$.

The AC current expression above is exact for the featureless-bath model at finite coupling. 
The $k=0$ term does not contribute, implying zero time average.
The $k$-th harmonic is determined by overlaps between micromotion Fourier components separated by $k$, namely $[P_\xi^\dagger P_{\xi+k}]_{aa}$.
By contrast, the DC Floquet-Landauer formula depends on the diagonal sideband weights $[P_\xi^\dagger P_\xi]_{aa}$.

To obtain the weak-coupling limit, we take $\Gamma_L,\Gamma_R \to 0^+$, with their ratio fixed. 
For every nonzero integer $k$, $\frac{k\hbar\Omega}{k\hbar\Omega+i\Gamma} \to 1$, while
$f_{L,\Gamma/2}(E) \to f_L(E)$.
The AC contribution is therefore first order in the left reservoir coupling.
This is of the same order as the DC current in Eq.~\eqref{eq:floquet_landauer}, explaining why their magnitudes can be comparable even in the weak-coupling regime.
We clarify that the designation ``AC'' refers to the contribution defined in Eq.~\eqref{eq:ac_current}.
Although its time average vanishes exactly, the ``DC'' terms of Eq.~\eqref{eq:kadanoff_baym_left} may also contain oscillatory contributions at finite coupling. 
The complete time-dependent current includes both contributions.

\bibliography{references}

@article{bukov_universal_2015,
  title = {Universal High-Frequency Behavior of Periodically Driven Systems: From Dynamical Stabilization to {{Floquet}} Engineering},
  shorttitle = {Universal High-Frequency Behavior of Periodically Driven Systems},
  author = {Bukov, Marin and D'Alessio, Luca and Polkovnikov, Anatoli},
  year = 2015,
  month = mar,
  journal = {Advances in Physics},
  volume = {64},
  number = {2},
  pages = {139--226},
  publisher = {Taylor \& Francis},
  issn = {0001-8732},
  doi = {10.1080/00018732.2015.1055918}
}

@article{bandyopadhyay_floquet_2022,
  title = {Floquet Engineering of {{Lie}} Algebraic Quantum Systems},
  author = {Bandyopadhyay, Jayendra N. and Thingna, Juzar},
  year = 2022,
  month = jan,
  journal = {Physical Review B},
  volume = {105},
  number = {2},
  pages = {L020301},
  publisher = {American Physical Society},
  doi = {10.1103/PhysRevB.105.L020301}
}

@article{tsuji_correlated_2008,
  title = {Correlated Electron Systems Periodically Driven out of Equilibrium: {{Floquet+DMFT}} Formalism},
  shorttitle = {Correlated Electron Systems Periodically Driven out of Equilibrium},
  author = {Tsuji, Naoto and Oka, Takashi and Aoki, Hideo},
  year = 2008,
  month = dec,
  journal = {Physical Review B},
  volume = {78},
  number = {23},
  pages = {235124},
  publisher = {American Physical Society},
  doi = {10.1103/PhysRevB.78.235124}
}

@article{pan_asymmetry-induced_2025,
  title = {Asymmetry-Induced Radiative Heat Transfer in {{Floquet}} Systems},
  author = {Pan, Hui and Ren, Yuhua and Tang, Gaomin and Wang, Jian-Sheng},
  year = 2025,
  month = jul,
  journal = {Physical Review B},
  volume = {112},
  number = {4},
  pages = {L041401},
  publisher = {American Physical Society},
  doi = {10.1103/74rq-f642}
}

@misc{rudner_floquet_2020,
  title = {The {{Floquet}} Engineer's Handbook},
  author = {Rudner, Mark S. and Lindner, Netanel H.},
  year = 2020,
  month = jun,
  number = {arXiv:2003.08252},
  eprint = {2003.08252},
  primaryclass = {cond-mat},
  publisher = {arXiv},
  doi = {10.48550/arXiv.2003.08252},
  urldate = {2026-03-05},
  archiveprefix = {arXiv}
}

@article{shirley_solution_1965,
  title = {Solution of the {{Schr\"odinger}} equation with a {{Hamiltonian}} periodic in time},
  author = {Shirley, Jon H.},
  year = 1965,
  month = may,
  journal = {Physical Review},
  volume = {138},
  number = {4B},
  pages = {B979-B987},
  publisher = {American Physical Society},
  doi = {10.1103/PhysRev.138.B979}
}

@article{matsyshyn_fermi-dirac_2023,
  title = {Fermi-{{Dirac}} Staircase Occupation of {{Floquet}} Bands and Current Rectification inside the Optical Gap of Metals: An Exact Approach},
  shorttitle = {Fermi-{{Dirac}} Staircase Occupation of {{Floquet}} Bands and Current Rectification inside the Optical Gap of Metals},
  author = {Matsyshyn, Oles and Song, Justin C. W. and Villadiego, Inti Sodemann and Shi, Li-kun},
  year = 2023,
  month = may,
  journal = {Physical Review B},
  volume = {107},
  number = {19},
  pages = {195135},
  publisher = {American Physical Society},
  doi = {10.1103/PhysRevB.107.195135}
}

@article{shi_floquet_2024,
  title = {{{Floquet Fermi}} Liquid},
  author = {Shi, Li-kun and Matsyshyn, Oles and Song, Justin C. W. and Villadiego, Inti Sodemann},
  year = 2024,
  month = apr,
  journal = {Physical Review Letters},
  volume = {132},
  number = {14},
  pages = {146402},
  publisher = {American Physical Society},
  doi = {10.1103/PhysRevLett.132.146402}
}

@article{tang_modulating_2024,
  title = {Modulating Near-Field Thermal Transfer through Temporal Drivings: {{A}} Quantum Many-Body Theory},
  shorttitle = {Modulating Near-Field Thermal Transfer through Temporal Drivings},
  author = {Tang, Gaomin and Wang, Jian-Sheng},
  year = 2024,
  month = feb,
  journal = {Physical Review B},
  volume = {109},
  number = {8},
  pages = {085428},
  publisher = {American Physical Society},
  doi = {10.1103/PhysRevB.109.085428}
}

@article{wang_transport_2023,
  title = {Transport in Electron-Photon Systems},
  author = {Wang, Jian-Sheng and Peng, Jiebin and Zhang, Zu-Quan and Zhang, Yong-Mei and Zhu, Tao},
  year = {2023},
  month = mar,
  journal = {Frontiers of Physics},
  volume = {18},
  number = {4},
  pages = {43602},
  issn = {2095-0470},
  doi = {10.1007/s11467-023-1260-z},
  urldate = {2024-02-21}
}

@book{haug_quantum_2008,
  title={Quantum Kinetics in Transport and Optics of Semiconductors},
  author={Haug, Hartmut and Jauho, Antti-Pekka},
  edition={2},
  publisher={Springer}, 
  address={Berlin},
  year={2008}
}

@book{stefanucci_nonequilibrium_2013, 
  title={Nonequilibrium Many-Body Theory of Quantum Systems: A Modern Introduction}, 
  author={Stefanucci, Gianluca and van Leeuwen, Robert}, 
  publisher={Cambridge University Press}, 
  address={Cambridge}, 
  year={2013}
}

@book{datta_mesoscopic_1995,
  author    = {Supriyo Datta},
  title     = {Electronic Transport in Mesoscopic Systems},
  publisher = {Cambridge University Press},
  year      = {1995},
  series    = {Cambridge Studies in Semiconductor Physics and Microelectronic Engineering},
  isbn      = {0521416043},
  doi       = {10.1017/CBO9780511805776}
}

@article{kohler_floquet-markovian_1997,
  title = {Floquet-{{Markovian}} Description of the Parametrically Driven, Dissipative Harmonic Quantum Oscillator},
  author = {Kohler, Sigmund and Dittrich, Thomas and H{\"a}nggi, Peter},
  year = 1997,
  month = jan,
  journal = {Physical Review E},
  volume = {55},
  number = {1},
  pages = {300--313},
  publisher = {American Physical Society},
  doi = {10.1103/PhysRevE.55.300}
}

@article{kohn_periodic_2001,
  title = {Periodic Thermodynamics},
  author = {Kohn, Walter},
  year = 2001,
  month = may,
  journal = {Journal of Statistical Physics},
  volume = {103},
  number = {3},
  pages = {417--423},
  issn = {1572-9613},
  doi = {10.1023/A:1010327828445}
}

@article{hone_statistical_2009,
  title = {Statistical Mechanics of {{Floquet}} Systems: The Pervasive Problem of near Degeneracies},
  shorttitle = {Statistical Mechanics of {{Floquet}} Systems},
  author = {Hone, Daniel W. and Ketzmerick, Roland and Kohn, Walter},
  year = 2009,
  month = may,
  journal = {Physical Review E},
  volume = {79},
  number = {5},
  pages = {051129},
  publisher = {American Physical Society},
  doi = {10.1103/PhysRevE.79.051129}
}

@article{ketzmerick_statistical_2010,
  title = {Statistical Mechanics of {{Floquet}} Systems with Regular and Chaotic States},
  author = {Ketzmerick, Roland and Wustmann, Waltraut},
  year = 2010,
  month = aug,
  journal = {Physical Review E},
  volume = {82},
  number = {2},
  pages = {021114},
  publisher = {American Physical Society},
  doi = {10.1103/PhysRevE.82.021114}
}

@article{iadecola_floquet_2015,
  title = {{{Floquet}} Systems Coupled to Particle Reservoirs},
  author = {Iadecola, Thomas and Chamon, Claudio},
  year = 2015,
  month = may,
  journal = {Physical Review B},
  volume = {91},
  number = {18},
  pages = {184301},
  publisher = {American Physical Society},
  doi = {10.1103/PhysRevB.91.184301}
}

@article{shirai_condition_2015,
  title = {Condition for Emergence of the {{Floquet-Gibbs}} State in Periodically Driven Open Systems},
  author = {Shirai, Tatsuhiko and Mori, Takashi and Miyashita, Seiji},
  year = 2015,
  month = mar,
  journal = {Physical Review E},
  volume = {91},
  number = {3},
  pages = {030101},
  publisher = {American Physical Society},
  doi = {10.1103/PhysRevE.91.030101}
}

@article{shirai_effective_2016,
  title = {Effective {{Floquet}}--{{Gibbs}} States for Dissipative Quantum Systems},
  author = {Shirai, Tatsuhiko and Thingna, Juzar and Mori, Takashi and Denisov, Sergey and H{\"a}nggi, Peter and Miyashita, Seiji},
  year = 2016,
  month = may,
  journal = {New Journal of Physics},
  volume = {18},
  number = {5},
  pages = {053008},
  publisher = {IOP Publishing},
  issn = {1367-2630},
  doi = {10.1088/1367-2630/18/5/053008}
}

@article{campaioli_quantum_2024,
  title = {Quantum Master Equations: Tips and Tricks for Quantum Optics, Quantum Computing, and Beyond},
  shorttitle = {Quantum Master Equations},
  author = {Campaioli, Francesco and Cole, Jared H. and Hapuarachchi, Harini},
  year = 2024,
  month = jun,
  journal = {PRX Quantum},
  volume = {5},
  number = {2},
  pages = {020202},
  publisher = {American Physical Society},
  doi = {10.1103/PRXQuantum.5.020202}
}

@article{blanes_magnus_2009,
  title = {The {{Magnus}} Expansion and Some of Its Applications},
  author = {Blanes, S. and Casas, F. and Oteo, J. A. and Ros, J.},
  year = 2009,
  month = jan,
  journal = {Physics Reports},
  volume = {470},
  number = {5},
  pages = {151--238},
  issn = {0370-1573},
  doi = {10.1016/j.physrep.2008.11.001}
}

@article{leggett_dynamics_1987,
  title = {Dynamics of the Dissipative Two-State System},
  author = {Leggett, A. J. and Chakravarty, S. and Dorsey, A. T. and Fisher, Matthew P. A. and Garg, Anupam and Zwerger, W.},
  year = 1987,
  month = jan,
  journal = {Reviews of Modern Physics},
  volume = {59},
  number = {1},
  pages = {1--85},
  publisher = {American Physical Society},
  doi = {10.1103/RevModPhys.59.1}
}

@article{haughian_quantum_2018,
  title = {Quantum Thermodynamics of the Resonant-Level Model with Driven System-Bath Coupling},
  author = {Haughian, Patrick and Esposito, Massimiliano and Schmidt, Thomas L.},
  year = 2018,
  month = feb,
  journal = {Physical Review B},
  volume = {97},
  number = {8},
  pages = {085435},
  publisher = {American Physical Society},
  doi = {10.1103/PhysRevB.97.085435}
}

@article{raju_quantum_2016,
  title = {Quantum Dissipative Effects on Non-Equilibrium Transport through a Single-Molecular Transistor: {{The Anderson-Holstein-Caldeira-Leggett}} Model},
  shorttitle = {Quantum Dissipative Effects on Non-Equilibrium Transport through a Single-Molecular Transistor},
  author = {Raju, Ch Narasimha and Chatterjee, Ashok},
  year = 2016,
  month = jan,
  journal = {Scientific Reports},
  volume = {6},
  number = {1},
  pages = {18511},
  publisher = {Nature Publishing Group},
  issn = {2045-2322},
  doi = {10.1038/srep18511}
}

@article{thakur_tutorial_2023,
  title = {A Tutorial on the {{NEGF}} Method for Electron Transport in Devices and Defective Materials},
  author = {Thakur, Akansha and Sarkar, Niladri},
  year = 2023,
  month = aug,
  journal = {The European Physical Journal B},
  volume = {96},
  number = {8},
  pages = {113},
  issn = {1434-6036},
  doi = {10.1140/epjb/s10051-023-00580-5}
}

@article{camalet_current_2003,
  title = {Current Noise in Ac-Driven Nanoscale Conductors},
  author = {Camalet, S{\'e}bastien and Lehmann, J{\"o}rg and Kohler, Sigmund and H{\"a}nggi, Peter},
  year = 2003,
  month = may,
  journal = {Physical Review Letters},
  volume = {90},
  number = {21},
  pages = {210602},
  publisher = {American Physical Society},
  doi = {10.1103/PhysRevLett.90.210602}
}

@article{wu_floquetgreens_2008,
  title = {A {{Floquet}}--{{Green}}'s Function Approach to Mesoscopic Transport under Ac Bias},
  author = {Wu, B H and Cao, J C},
  year = 2008,
  month = feb,
  journal = {Journal of Physics: Condensed Matter},
  volume = {20},
  number = {8},
  pages = {085224},
  issn = {0953-8984},
  doi = {10.1088/0953-8984/20/8/085224},
}

@article{stefanucci_time-dependent_2008,
  title = {Time-Dependent Approach to Electron Pumping in Open Quantum Systems},
  author = {Stefanucci, G. and Kurth, S. and Rubio, A. and Gross, E. K. U.},
  year = 2008,
  month = feb,
  journal = {Physical Review B},
  volume = {77},
  number = {7},
  pages = {075339},
  publisher = {American Physical Society},
  doi = {10.1103/PhysRevB.77.075339}
}

@article{mosallanejad_floquet_2024,
  title = {Floquet Non-Equilibrium {{Green}}'s Function and {{Floquet}} Quantum Master Equation for Electronic Transport: {{The}} Role of Electron--Electron Interactions and Spin Current with Circular Light},
  shorttitle = {Floquet Non-Equilibrium {{Green}}'s Function and {{Floquet}} Quantum Master Equation for Electronic Transport},
  author = {Mosallanejad, Vahid and Wang, Yu and Dou, Wenjie},
  year = 2024,
  month = apr,
  journal = {The Journal of Chemical Physics},
  volume = {160},
  number = {16},
  pages = {164102},
  issn = {0021-9606},
  doi = {10.1063/5.0184978}
}

@article{caroli_direct_1971,
  title = {Direct Calculation of the Tunneling Current},
  author = {Caroli, C. and Combescot, R. and Nozieres, P. and {Saint-James}, D.},
  year = 1971,
  month = jun,
  journal = {Journal of Physics C: Solid State Physics},
  volume = {4},
  number = {8},
  pages = {916},
  issn = {0022-3719},
  doi = {10.1088/0022-3719/4/8/018}
}

@article{lindner_floquet_2011,
  title = {Floquet Topological Insulator in Semiconductor Quantum Wells},
  author = {Lindner, Netanel H. and Refael, Gil and Galitski, Victor},
  year = 2011,
  month = jun,
  journal = {Nature Physics},
  volume = {7},
  number = {6},
  pages = {490--495},
  publisher = {Nature Publishing Group},
  issn = {1745-2481},
  doi = {10.1038/nphys1926}
}

@article{oka_photovoltaic_2009,
  title = {Photovoltaic {{Hall}} Effect in Graphene},
  author = {Oka, Takashi and Aoki, Hideo},
  year = 2009,
  month = feb,
  journal = {Physical Review B},
  volume = {79},
  number = {8},
  pages = {081406},
  publisher = {American Physical Society},
  doi = {10.1103/PhysRevB.79.081406}
}

@article{wang_observation_2013,
  title = {Observation of {{Floquet-Bloch States}} on the {{Surface}} of a {{Topological Insulator}}},
  author = {Wang, Y. H. and Steinberg, H. and {Jarillo-Herrero}, P. and Gedik, N.},
  year = 2013,
  month = oct,
  journal = {Science},
  volume = {342},
  number = {6157},
  pages = {453--457},
  publisher = {American Association for the Advancement of Science},
  doi = {10.1126/science.1239834}
}

@article{wang_floquet_2024,
  title = {Floquet Engineering Tunable Periodic Gauge Fields and Simulating Real Topological Phases in a Cold-Alkaline-Earth-Metal-Atom Optical Lattice},
  author = {Wang, Wei and Zhang, Zheng and Tang, Gui-Xin and Wang, Tao},
  year = 2024,
  month = aug,
  journal = {Physical Review A},
  volume = {110},
  number = {2},
  pages = {023308},
  publisher = {American Physical Society},
  doi = {10.1103/PhysRevA.110.023308}
}

@article{jiao_floquet_2026,
  title = {Floquet Control of Modulational Instability in a Spin-Orbit Coupled Condensate},
  author = {Jiao, Chen and Bai, Wen-Kai and Zheng, Jun-Hui and Yang, Tao},
  year = 2026,
  month = jul,
  journal = {Chaos, Solitons \& Fractals},
  volume = {208},
  pages = {118172},
  issn = {0960-0779},
  doi = {10.1016/j.chaos.2026.118172}
}

@article{rudner_band_2020,
  title = {Band Structure Engineering and Non-Equilibrium Dynamics in {{Floquet}} Topological Insulators},
  author = {Rudner, Mark S. and Lindner, Netanel H.},
  year = 2020,
  month = may,
  journal = {Nature Reviews Physics},
  volume = {2},
  number = {5},
  pages = {229--244},
  publisher = {Nature Publishing Group},
  issn = {2522-5820},
  doi = {10.1038/s42254-020-0170-z}
}

@article{castro_floquet_2022,
  title = {Floquet Engineering the Band Structure of Materials with Optimal Control Theory},
  author = {Castro, Alberto and De Giovannini, Umberto and Sato, Shunsuke A. and H{\"u}bener, Hannes and Rubio, Angel},
  year = 2022,
  month = sep,
  journal = {Physical Review Research},
  volume = {4},
  number = {3},
  pages = {033213},
  publisher = {American Physical Society},
  doi = {10.1103/PhysRevResearch.4.033213}
}

@article{sie_valley-selective_2015,
  title = {Valley-Selective Optical {{Stark}} Effect in Monolayer {{WS2}}},
  author = {Sie, Edbert J. and McIver, James W. and Lee, Yi-Hsien and Fu, Liang and Kong, Jing and Gedik, Nuh},
  year = 2015,
  month = mar,
  journal = {Nature Materials},
  volume = {14},
  number = {3},
  pages = {290--294},
  publisher = {Nature Publishing Group},
  issn = {1476-4660},
  doi = {10.1038/nmat4156}
}

@article{esin_floquet_2020,
  title = {Floquet Metal-to-Insulator Phase Transitions in Semiconductor Nanowires},
  author = {Esin, Iliya and Rudner, Mark S. and Lindner, Netanel H.},
  year = 2020,
  month = aug,
  journal = {Science Advances},
  volume = {6},
  number = {35},
  pages = {eaay4922},
  publisher = {American Association for the Advancement of Science},
  doi = {10.1126/sciadv.aay4922}
}

@article{deng_observation_2015,
  title = {Observation of {{Floquet States}} in a {{Strongly Driven Artificial Atom}}},
  author = {Deng, Chunqing and Orgiazzi, Jean-Luc and Shen, Feiruo and Ashhab, Sahel and Lupascu, Adrian},
  year = 2015,
  month = sep,
  journal = {Physical Review Letters},
  volume = {115},
  number = {13},
  pages = {133601},
  publisher = {American Physical Society},
  doi = {10.1103/PhysRevLett.115.133601}
}

@article{zhao_probing_2022,
  title = {Probing {{Operator Spreading}} via {{Floquet Engineering}} in a {{Superconducting Circuit}}},
  author = {Zhao, S. K. and Ge, Zi-Yong and Xiang, Zhongcheng and Xue, G. M. and Yan, H. S. and Wang, Z. T. and Wang, Zhan and Xu, H. K. and Su, F. F. and Yang, Z. H. and Zhang, He and Zhang, Yu-Ran and Guo, Xue-Yi and Xu, Kai and Tian, Ye and Yu, H. F. and Zheng, D. N. and Fan, Heng and Zhao, S. P.},
  year = 2022,
  month = oct,
  journal = {Physical Review Letters},
  volume = {129},
  number = {16},
  pages = {160602},
  publisher = {American Physical Society},
  doi = {10.1103/PhysRevLett.129.160602}
}

@article{jotzu_experimental_2014,
  title = {Experimental Realization of the Topological {{Haldane}} Model with Ultracold Fermions},
  author = {Jotzu, Gregor and Messer, Michael and Desbuquois, R{\'e}mi and Lebrat, Martin and Uehlinger, Thomas and Greif, Daniel and Esslinger, Tilman},
  year = 2014,
  month = nov,
  journal = {Nature},
  volume = {515},
  number = {7526},
  pages = {237--240},
  publisher = {Nature Publishing Group},
  issn = {1476-4687},
  doi = {10.1038/nature13915}
}

@article{meinert_floquet_2016,
  title = {Floquet {{Engineering}} of {{Correlated Tunneling}} in the {{Bose-Hubbard Model}} with {{Ultracold Atoms}}},
  author = {Meinert, F. and Mark, M. J. and Lauber, K. and Daley, A. J. and N{\"a}gerl, H.-C.},
  year = 2016,
  month = may,
  journal = {Physical Review Letters},
  volume = {116},
  number = {20},
  pages = {205301},
  publisher = {American Physical Society},
  doi = {10.1103/PhysRevLett.116.205301}
}

@article{miller_two-axis_2024,
  title = {Two-Axis Twisting Using {{Floquet-engineered XYZ}} Spin Models with Polar Molecules},
  author = {Miller, Calder and Carroll, Annette N. and Lin, Junyu and Hirzler, Henrik and Gao, Haoyang and Zhou, Hengyun and Lukin, Mikhail D. and Ye, Jun},
  year = 2024,
  month = sep,
  journal = {Nature},
  volume = {633},
  number = {8029},
  pages = {332--337},
  publisher = {Nature Publishing Group},
  issn = {1476-4687},
  doi = {10.1038/s41586-024-07883-2}
}

@article{rechtsman_photonic_2013,
  title = {Photonic {{Floquet}} Topological Insulators},
  author = {Rechtsman, Mikael C. and Zeuner, Julia M. and Plotnik, Yonatan and Lumer, Yaakov and Podolsky, Daniel and Dreisow, Felix and Nolte, Stefan and Segev, Mordechai and Szameit, Alexander},
  year = 2013,
  month = apr,
  journal = {Nature},
  volume = {496},
  number = {7444},
  pages = {196--200},
  publisher = {Nature Publishing Group},
  issn = {1476-4687},
  doi = {10.1038/nature12066}
}

@article{maczewsky_observation_2017,
  title = {Observation of Photonic Anomalous {{Floquet}} Topological Insulators},
  author = {Maczewsky, Lukas J. and Zeuner, Julia M. and Nolte, Stefan and Szameit, Alexander},
  year = 2017,
  month = jan,
  journal = {Nature Communications},
  volume = {8},
  number = {1},
  pages = {13756},
  publisher = {Nature Publishing Group},
  issn = {2041-1723},
  doi = {10.1038/ncomms13756}
}

@article{arrachea_green-function_2005,
  title = {Green-Function Approach to Transport Phenomena in Quantum Pumps},
  author = {Arrachea, Liliana},
  year = 2005,
  month = sep,
  journal = {Physical Review B},
  volume = {72},
  number = {12},
  pages = {125349},
  publisher = {American Physical Society},
  doi = {10.1103/PhysRevB.72.125349}
}

@article{breuer_quasistationary_2000,
  title = {Quasistationary Distributions of Dissipative Nonlinear Quantum Oscillators in Strong Periodic Driving Fields},
  author = {Breuer, Heinz-Peter and Huber, Wolfgang and Petruccione, Francesco},
  year = 2000,
  month = may,
  journal = {Physical Review E},
  volume = {61},
  number = {5},
  pages = {4883--4889},
  publisher = {American Physical Society},
  doi = {10.1103/PhysRevE.61.4883}
}

@article{dittrich_driven_1993,
  title = {Driven Tunnelling with Dissipation},
  author = {Dittrich, T. and Oelschl{\"a}gel, B. and H{\"a}nggi, P.},
  year = 1993,
  month = apr,
  journal = {Europhysics Letters},
  volume = {22},
  number = {1},
  pages = {5},
  issn = {0295-5075},
  doi = {10.1209/0295-5075/22/1/002}
}

@article{liu_classification_2015,
  title = {Classification of the {{Floquet}} Statistical Distribution for Time-Periodic Open Systems},
  author = {Liu, Dong E.},
  year = 2015,
  month = apr,
  journal = {Physical Review B},
  volume = {91},
  number = {14},
  pages = {144301},
  publisher = {American Physical Society},
  doi = {10.1103/PhysRevB.91.144301}
}

@article{liu_keldysh_2017,
  title = {Keldysh Approach to Periodically Driven Systems with a Fermionic Bath: Nonequilibrium Steady State, Proximity Effect, and Dissipation},
  shorttitle = {Keldysh Approach to Periodically Driven Systems with a Fermionic Bath},
  author = {Liu, Dong E. and Levchenko, Alex and Lutchyn, Roman M.},
  year = 2017,
  month = mar,
  journal = {Physical Review B},
  volume = {95},
  number = {11},
  pages = {115303},
  publisher = {American Physical Society},
  doi = {10.1103/PhysRevB.95.115303}
}

@article{stafford_resonant_1996,
  title = {Resonant Photon-Assisted Tunneling through a Double Quantum Dot: An Electron Pump from Spatial {{Rabi}} Oscillations},
  shorttitle = {Resonant Photon-Assisted Tunneling through a Double Quantum Dot},
  author = {Stafford, C. A. and Wingreen, Ned S.},
  year = 1996,
  month = mar,
  journal = {Physical Review Letters},
  volume = {76},
  number = {11},
  pages = {1916--1919},
  publisher = {American Physical Society},
  doi = {10.1103/PhysRevLett.76.1916}
}

@article{peralta_gavensky_stifmmode_2025,
  title = {{{St\v{r}eda}} Formula for {{Floquet}} Systems: Topological Invariants and Quantized Anomalies from {{Ces\`aro}} Summation},
  shorttitle = {{{St\v{r}eda}} Formula for {{Floquet}} Systems},
  author = {Peralta Gavensky, Lucila and Usaj, Gonzalo and Goldman, Nathan},
  year = 2025,
  month = sep,
  journal = {Physical Review X},
  volume = {15},
  number = {3},
  pages = {031067},
  publisher = {American Physical Society},
  doi = {10.1103/b3pw-my97}
}

@article{kumari_josephson-current_2024,
  title = {{{Josephson}}-Current Signatures of Unpaired {{Floquet}} {{Majorana}} Fermions},
  author = {Kumari, Rekha and Seradjeh, Babak and Kundu, Arijit},
  year = 2024,
  month = nov,
  journal = {Physical Review Letters},
  volume = {133},
  number = {19},
  pages = {196601},
  publisher = {American Physical Society},
  doi = {10.1103/PhysRevLett.133.196601}
}

@article{ridley_calculation_2016,
  title = {Calculation of the Current Response in a Nanojunction for an Arbitrary Time-Dependent Bias: Application to the Molecular Wire},
  shorttitle = {Calculation of the Current Response in a Nanojunction for an Arbitrary Time-Dependent Bias},
  author = {Ridley, Michael and MacKinnon, Angus and Kantorovich, Lev},
  year = 2016,
  month = mar,
  journal = {Journal of Physics: Conference Series},
  volume = {696},
  number = {1},
  pages = {012017},
  publisher = {IOP Publishing},
  issn = {1742-6596},
  doi = {10.1088/1742-6596/696/1/012017}
}

\end{document}